\documentclass[conference]{IEEEtran}
\usepackage{cite}
\usepackage{amsmath,amssymb,amsfonts}
\usepackage{algorithmic}
\usepackage{graphicx}
\usepackage{textcomp}
\usepackage{xcolor}
\usepackage{enumitem}
\usepackage{balance}

\usepackage{amsmath,amssymb,amsfonts}

\usepackage{todonotes}
\usepackage{xcolor} 
\usepackage{xspace} 
\usepackage{url}

\usepackage{graphicx}

\usepackage{tikz}
\usetikzlibrary{shapes,patterns}

\usepackage{pgfplots}
\usepackage{pgfplotstable}
\usepackage{tikzscale}
\pgfplotsset{compat=1.15}

\usepackage{algorithm}

\usepackage{gnuplottex}

\usepackage{xfrac}
\usepackage{blindtext}

\usepackage{relsize} 

\usepackage{subcaption}
\definecolor{cb1_black}{cmyk}{0,0,0,1}              
\definecolor{cb1_orange}{cmyk}{0,0.5,1,0}           
\definecolor{cb1_skyblue}{cmyk}{0.8,0,0,0}          
\definecolor{cb1_bluishgreen}{cmyk}{0.97,0,9.75,0}  
\definecolor{cb1_yellow}{cmyk}{0.1,0.05,0.9,0}      
\definecolor{cb1_blue}{cmyk}{1,0.5,0,0}             
\definecolor{cb1_vermillion}{cmyk}{0,0.8,1,0}       
\definecolor{cb1_reddishpurple}{cmyk}{0.1,0.7,0,0}  

\colorlet{black}{cb1_black}
\colorlet{orange}{cb1_orange}
\colorlet{lightblue}{cb1_skyblue}
\colorlet{green}{cb1_bluishgreen}
\colorlet{yellow}{cb1_yellow}
\colorlet{blue}{cb1_blue}
\colorlet{red}{cb1_vermillion}
\colorlet{purple}{cb1_reddishpurple}

\pgfplotscreateplotcyclelist{mylist}{
  {orange,mark=*},
  {blue,mark=square},
  {green,mark=o},
  {purple,mark=+},
  {lightblue,mark options={fill=brown!40},mark=otimes*},
  {red,mark options={fill=brown!40},mark=otimes*},
  {yellow,mark options={fill=brown!40},mark=otimes*},
}

\newcommand{\sysname}{\textsf{CLASH}\xspace}

\DeclareMathOperator*{\bigjoin}{\mathlarger{\Join}}

\newcommand{\se}[1]{{\operatorname{\mathit{#1}}}}

\makeatletter
\newcommand*{\rom}[1]{\expandafter\@slowromancap\romannumeral #1@}
\makeatother

\newcommand*\circled[2][white]{\tikz[baseline=(char.base)]{
        \node[shape=circle,draw,fill=#1,fill opacity=0.2, text opacity=1, inner sep=1pt] (char) {\footnotesize #2};}}
\newcommand*\patterncircled[2][white]{\tikz[baseline=(char.base)]{
        \node[shape=circle,draw,pattern color=#1,fill opacity=0.2, text opacity=1, inner sep=1pt,pattern=north west lines] (char) {\footnotesize #2};}}

\begin{document}


\pagestyle{plain}
\pagenumbering{arabic}
\title{Optimizing Multiple Multi-Way Stream Joins}

\author{\IEEEauthorblockN{{Manuel Dossinger}\\
    \textit{TU Kaiserslautern (TUK)}\\
    Kaiserslautern, Germany \\
    dossinger@cs.uni-kl.de}
  \and
  \IEEEauthorblockN{{Sebastian Michel}\\
    \textit{TU Kaiserslautern (TUK)}\\
    Kaiserslautern, Germany \\
    michel@cs.uni-kl.de}
}

\maketitle


\begin{abstract}
We address the joint optimization of multiple stream joins in a scale-out architecture by tailoring prior work on multi-way stream joins to predicate-driven data partitioning schemes.
We present an integer linear programming (ILP) formulation for selecting the partitioning and tuple routing with minimal probe load and describe how routing and operator placement can be rewired dynamically at changing data characteristics and arrival or expiration of queries.
The presented algorithms and optimization schemes are implemented in CLASH, a data stream processor developed in our group that translates queries to deployable Apache Storm topologies after optimization. The experiments conducted over real-world data exhibit the potential of multi-query optimization of multi-way stream joins and the effectiveness and feasibility of the ILP optimization problem.
\end{abstract}


\section{Introduction}\label{sec:introduction}

Processing data streams is a ubiquitous problem in data management, especially
in the context of handing large-volume streams in scale-out systems. Prominent engines like
Spark Streaming~\cite{website:spark}, Flink~\cite{DBLP:journals/debu/CarboneKEMHT15,website:flink}, Apache Storm's Trident~\cite{website:storm}, or Kafka~\cite{DBLP:journals/pvldb/WangKSPZNRKS15,website:kafka}
allow users or higher-level applications to express queries in SQL-style languages, deployed and
executed over potentially very many compute tasks in a data center. In most cases, such queries
do not merely filter or aggregate tuples from a single relation but involve {\it joins} that connect information pieces from various sources.
For instance, search-engine queries and ad-clicks need to be joined for billing purposes~\cite{DBLP:conf/sigmod/AnanthanarayananBDGJQRRSV13}
and in complex event processing, events are commonly expressed by multiple criteria that do not originate from a single sensor~\cite{DBLP:journals/pvldb/KolchinskyS18}.
Once such queries are posted, they remain registered in the system and continuously report results.
Computational resources can be shared for answering multiple queries.
Trying to share work between data stream queries is not a new idea~\cite{DBLP:conf/vldb/HammadFAE03,DBLP:conf/dasfaa/YangZWX19,DBLP:conf/sigmod/KarimovRM19},
and traditionally, sharing is done on a per-operator basis.
This means, a query execution plan is produced for each query and joins or other operators share their result with downstream operators to contribute to answer multiple queries at once.

In this paper, we propose a fine-grained and easy to rewire tuple routing scheme that enables sharing of partial results between multiple multi-way join queries.
Our approach is implemented in \sysname\footnote{https://www.youtube.com/watch?v=oZxNIwvEQDw}~\cite{DBLP:conf/sigmod/DossingerMR19}, a system for optimizing continuous queries, built on top of Apache Storm~\cite{DBLP:conf/sigmod/ToshniwalTSRPKJGFDBMR14,website:storm}.
However, the concepts are applicable to all modern streaming systems that give control over the routing between workers and access to local state.

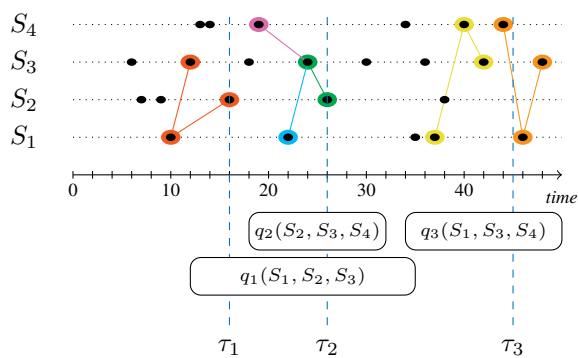
\begin{figure}[tb]
  \centering
  \begin{tikzpicture}[xscale=1.3]

    \draw[->] (0,0) to (5,0);

    \foreach \x in {0,0.2,...,4.8} {
        \draw (\x,0.05) -- (\x,-0.05);
      }

    \node[] at (0,-0.2){\scriptsize 0};
    \node[] at (1,-0.2){\scriptsize 10};
    \node[] at (2,-0.2){\scriptsize 20};
    \node[] at (3,-0.2){\scriptsize 30};
    \node[] at (4,-0.2){\scriptsize 40};
    \node[] at (5,-0.3){\it \scriptsize time};


    \node[] at (-.5,0.5){$S_1$};
    \node[] at (-.5,1.0){$S_2$};
    \node[] at (-.5,1.5){$S_3$};
    \node[] at (-.5,2.0){$S_4$};

    \draw[-, dotted] (0,0.5) to (5,0.5);
    \draw[-, dotted] (0,1.0) to (5,1.0);
    \draw[-, dotted] (0,1.5) to (5,1.5);
    \draw[-, dotted] (0,2.0) to (5,2.0);

    \fill[color = red] (1.6,1.0) circle (0.1);
    \fill[color = red] (1.2,1.5) circle (0.1);
    \fill[color = red] (1,0.5) circle (0.1);
    \draw[-, color=red] (1.6,1.0) to (1,0.5) to (1.2,1.5);

    \fill[color = green] (2.6,1.0) circle (0.1);
    \fill[color = green] (2.4,1.5) circle (0.1);
    \fill[color = lightblue] (2.2,.5) circle (0.1);
    \fill[color = purple] (1.9,2.0) circle (0.1);
    \draw[-, color=green] (2.6,1.0) to (2.4,1.5);
    \draw[-, color=lightblue] (2.4,1.5) to (2.2,.5);
    \draw[-, color=purple] (2.4,1.5) to (1.9,2.0);

    \fill[color = yellow] (4.2,1.5) circle (0.1);
    \fill[color = yellow] (4.0,2.0) circle (0.1);
    \fill[color = yellow] (3.7,0.5) circle (0.1);
    \draw[-, color=yellow] (4.2,1.5) to (4.0,2.0) to (3.7,0.5);

    \fill[color = orange] (4.8,1.5) circle (0.1);
    \fill[color = orange] (4.6,0.5) circle (0.1);
    \fill[color = orange] (4.4,2.0) circle (0.1);
    \draw[-, color=orange] (4.8,1.5) to (4.6,0.5) to (4.4,2.0);

    \foreach \x in {1,2.2,3.5,3.7,4.6} { \fill (\x,0.5) circle (0.05); } 
    \foreach \x in {.7,.9,1.6,2.6,3.8} { \fill (\x,1.0) circle (0.05); } 
    \foreach \x in {.6,1.2,1.8,2.4,3.0,3.6,4.2,4.8} { \fill (\x,1.5) circle (0.05); } 
    \foreach \x in {1.3,1.4,1.9,3.4,4.0,4.4} { \fill (\x,2.0) circle (0.05); } 


    \draw[blue,dashed] (1.6,-2) to (1.6,2.1); \node[] at (1.6,-2.3){$\tau_1$};
    \draw[blue,dashed] (2.6,-2) to (2.6,2.1); \node[] at (2.6,-2.3){$\tau_2$};
    \draw[blue,dashed] (4.5,-2) to (4.5,2.1); \node[] at (4.5,-2.3){$\tau_3$};


    \draw[rounded corners, fill=white] (1.2,-1.1) rectangle node {\scriptsize $q_1 (S_1, S_2, S_3)$} (3.5,-1.6);
    \draw[rounded corners, fill=white] (1.8,-.5) rectangle node {\scriptsize $q_2 (S_2, S_3, S_4)$} (3.2,-1);
    \draw[rounded corners, fill=white] (3.4,-.5) rectangle node {\scriptsize $q_3 (S_1, S_3, S_4)$} (5,-1);

  \end{tikzpicture}
  \caption{Four streaming relations with queries that are enabled and expire over time.}\label{fig:introexample}
\end{figure}

\subsection{Problem Statement}

In this paper, we consider optimizing multiple equi-join queries over streamed input relations $S_1, \dots, S_m$.
We write $S_i.a$ for naming an attribute, and $s_i.a$ for the value of attribute $a$ in tuple $s_i$. Tuples have a special attribute $\tau$ which is their timestamp.
The individual join predicates comprise predicates over pairs of relations, like $S_i.a = S_j.b$, where $a$ and $b$ are attributes of $S_i$ and $S_j$, respectively.
For each relation, a window defines the maximal time difference between a tuple of this relation and another tuple for being considered joinable.
When a tuple $s_i \in S_i$ is observed, immediately, join result tuples $s_1 \circ \dots \circ s_i \circ \dots \circ s_m$ are produced for all joinable tuples $s_1, \dots, s_{i-1}, s_{i+1}, \dots, s_m$ that arrived within the window depending on the arrival time of $s_i$.
For example, the tuples arriving at times 10, 12, and 16, marked in red in Figure~\ref{fig:introexample}, satisfy the join predicates and windows of query $q_1$.
Then the join result of this tuple is produced at time $\tau_1 = 16$, as soon as the tuple of $S_2$ arrives.

Given streamed relations $S_1, \dots, S_m$ and queries $q_1, \dots, q_k$ where each $q_i$ is a join query over a subset of the relations.
Then the goal is, to produce both, a selection of state to materialize on workers in a distributed environment, and a routing and probing strategy between these workers which produces the correct join result, has a low memory footprint and provides high throughput.
In order to enable efficient local join computation, cross products are avoided.
The processing scheme should adapt to changes in data characteristics and the addition of new or removal of old queries.

For example, in Figure~\ref{fig:introexample} at time $\tau_1$ query $q_1$ is known to the system, so a processing strategy optimal for this query is deployed.
This involves sending arriving tuples from $S_2$ to $S_1$ for producing an intermediate result and then sending this to $S_3$ for the complete result.
At time $\tau_2$ query $q_2$ is also incorporated into the system.
Now, tuples arriving from $S_2$ are routed differently: first to $S_3$ (marked green) in order to produce the partial join result, and then both to $S_4$ (marked purple) and to $S_1$ (marked blue) in order to produce both results for $q_2$ and $q_1$, respectively.
Both queries expire eventually and $q_3$ is installed initially with the probe strategy from $S_3$ to $S_4$ then to $S_1$.
At $\tau_3$ that data characteristics have changed and thus it is more efficient to sent $S_3$ tuples first to $S_1$ and then to $S_4$.

\subsection{Contribution and Outline}

In this paper, we make the following contributions:
\begin{enumerate}[label=\arabic*.]
  \item We provide a sound and feasible working solution for optimizing multiple multi-join queries in modern scale-out data stream processors.
  \item To do so, we provide an ILP formulation of data partitioning and probing optimization.
  \item We describe how on-the-fly rewiring of tuple routing can be realized to enable fast adaptation of shared plans and intra-plan tuple routing,
        for changing data characteristics as well as for arriving or expiring queries.
  \item We report on detailed finding of a thorough experimental evaluation and provide lessons learned in optimizing multiple multi-way join queries.
\end{enumerate}

The remainder of this paper is structured as follows. Section~\ref{sec:related_work} discusses related work
and Section~\ref{sec:preliminaries} briefly introduces fundamentals of stream join processing in scale-out environments.
Section~\ref{sec:distributing_stream_joins} describes the architectural setup and the core concepts of mating partitioning schemes imposed by equi-join predicates on tuple routing and storage.
Section~\ref{sec:join_optimization} presents the modeling of the optimization problem as integer linear program (ILP).
Section~\ref{sec:adaptive_probe_ordering} describes how tuple routing and result sharing can be adapted
to changing query load and data characteristics.
Section~\ref{sec:experiments} reports on the results of our experimental study and lessons learned.
Finally, Section~\ref{sec:conclusion} concludes the paper.


\section{Related Work}\label{sec:related_work}

There is ample research on the implementation of local joins (like the Grace hash join~\cite{DBLP:journals/ngc/KitsuregawaTM83} or the hybrid hash join~\cite{DBLP:conf/sigmod/DeWittKOSSW84}) as well as the optimization of join orders~\cite{DBLP:conf/sigmod/SelingerACLP79,DBLP:journals/vldb/SteinbrunnMK97}, and both topics are regularly revisited~\cite{DBLP:conf/sigmod/SchuhCD16,DBLP:journals/vldb/LeisRGMBKN18}.
Joglekar and R{\'e}~\cite{DBLP:conf/icdt/JoglekarR16} propose using information on the multiplicity of values to optimize multi-way joins, but not considering distributed computation (although some results are of generic nature).
Specifically addressing window stream joins (cf., \cite{DBLP:journals/tods/KramerS09}), Hammad et al.~\cite{DBLP:journals/vldb/HammadAE08} present two algorithms for processing multi-way joins in a centralized setting; there is no consideration of how such algorithms could potentially be executed in a distribution fashion.
The algorithms are, however, oblivious to the matching predicate, and, thus, not bound to equi joins.
Zhou et al.~\cite{DBLP:conf/dasfaa/ZhouYYZ06} propose an approach for minimizing the communication cost between nodes when evaluating a multi-way equi join.
Afrati et al.~\cite{DBLP:conf/icdt/AfratiJRSU17} present a multi-round algorithm with bounded communication cost for computing equi joins over multiple relations.

For data streaming applications, Viglas and Naughton~\cite{DBLP:conf/sigmod/ViglasN02} propose rate-based rather than classical cost-based optimization and in~\cite{DBLP:conf/vldb/ViglasNB03} they introduce the MJoin operator for multi-way join computation on a single node.
Golab and {\"O}zsu~\cite{DBLP:conf/vldb/GolabO03} process windowed stream joins on a single machine using multiple nested loop joins, where the join order is determined using the arrival rate of the streams and selectivity of the predicates.
Wang and Rundensteiner~\cite{DBLP:conf/edbt/WangR09} present a way to distribute the work of a single join operation over multiple stages by employing time-slicing of the join operators state.
Lin et al. present the BiStream~\cite{DBLP:conf/sigmod/LinOWY15} operator, which enables equi and theta joins over two relations without redundancy in state.

With partial key grouping, Nasir et al.~\cite{DBLP:journals/corr/NasirMGKS15a} introduce a value-based partitioning scheme that is able to reduce the load of individual nodes in a computing cluster if the partitioning key is skewed, and Qiu et al.~\cite{DBLP:conf/edbt/QiuPY19} apply streaming hypercube for heavily skewed data.

Madsen et al.~\cite{DBLP:conf/debs/MadsenZS16} discuss the benefits of exposing query intention to the system, rather than keeping it in the black box of UDFs.
Oguz et al.~\cite{DBLP:conf/adbis/OguzYHED16} propose changing the implementation during query answering from symmetric hash join to bind join and back, depending on arrival rates and result size.
Rödiger et al.~\cite{DBLP:conf/icde/RodigerIK016} and Li et al.~\cite{DBLP:conf/sigmod/LiRD18} split the handling of heavy hitters from the rest of the tuples.
Specifically for joins, Gomes et al.~\cite{DBLP:journals/inffus/GomesC08} propose changing roles of relations in a binary join tree.
Similarly, in DBMSs adaptive processing techniques are employed, e.g., for long running queries where the initially selected plan turns out to be suboptimal.
Li et al.~\cite{DBLP:conf/icde/LiSMBCL07} also changing the roles of relations in the join tree.
An example for tuple centric routing strategies for continuous query optimization is Avenur and Hellerstein's work on Eddies~\cite{DBLP:conf/sigmod/HellersteinA00} and later Distributed Eddies~\cite{DBLP:conf/vldb/TianD03}.
Query optimization using mathematical programming was explored by Trummer and Koch for join ordering~\cite{DBLP:conf/sigmod/Trummer017} and multi query optimization~\cite{DBLP:journals/pvldb/Trummer016}.
Dökeroğlu et al.~\cite{DBLP:conf/iscis/DokerogluBC14} propose multi query optimization using ILP for streaming data but only with binary joins.

Yang et al.~\cite{DBLP:conf/dasfaa/YangZWX19} propose cost-based optimization where operators are shared between plans for multiple queries if they produce the same or implied output streams.
Jonathan et al.~\cite{DBLP:conf/cloud/JonathanCW18} show multi-query optimization where multiple data centers are involved and slower inter-dc-communication is respected.
They introduce operator sharing strategies for saving both, computation and communication.
Kolchinsky and Schuster~\cite{DBLP:journals/pvldb/KolchinskyS18} propose optimization techniques for complex event processing systems where many patterns are registered simultaneously.
Karimov et al.~\cite{DBLP:conf/sigmod/KarimovRM19} present AStream, a system for sharing resources for multiple streaming queries.
They share parts of the history of joins however compared to our approach, only if exactly the same joins are used in different queries and they ignore partitioning.


\section{Scale-Out Stream Join Preliminaries}\label{sec:preliminaries}

Distributed stream joins are generally conducted as follows:
Tuples are placed on some compute nodes, and future tuples are routed to these compute nodes, such that all possible join partners meet and the correct result is produced.
A prominent example is the symmetric hash join~\cite{DBLP:journals/dpd/WilschutA93} where $R \Join S$ is computed by storing $R$-tuples in one set of nodes and $S$-tuples in another set of nodes.
In a scale-out architecture, multiple processes are responsible for storing subsets of individual relations.
For instance, there might be $n$ tasks instantiated to store the tuples of relation $R$.
When a tuple $s$ of relation $S$ arrives, dependent on the kind of join, it is sent to all or a dedicated subset of tasks of relation $R$.
Most importantly, all relevant tuples need to be stored as long as they might be join partners for incoming tuples.
Starting from this, there are several refinements, e.g., arranging the nodes in a hyper-cube scheme or doing a combination of random partitioning and broadcasting for computing theta joins~\cite{DBLP:journals/pvldb/VitorovicEGMEDK16,DBLP:conf/sigmod/LinOWY15}.

The described processing schemes allow trading off storage and communication cost.
    {\bf Storage cost} is the cost for deploying all partitions of input relations as well as additional intermediate relations.
The minimally required cost of this is achieved when only the input relations are materialized, and with each introduced intermediate store this cost grows.

    {\bf Communication cost} consists of two aspects: the size of all tuples sent across the computing topology and the number of messaging events.
A tuple $r \in R$ probing tuples of $S$ can produce between $0$ and $|S|$ many result tuples, increasing the number of tuples sent. However, the number of messages sent is always the same, as result tuples are sent together in one message.
We call the number of tuples sent the {\bf probe cost}, and this our subject of minimization.
Some systems opt to further group tuples into micro batches~\cite{website:spark} and thereby further reduce the number of messaging events.
However, the overall size of these messages still depends on the number of sent tuples.


\begin{figure*}[t]
  \includegraphics[width=\textwidth]{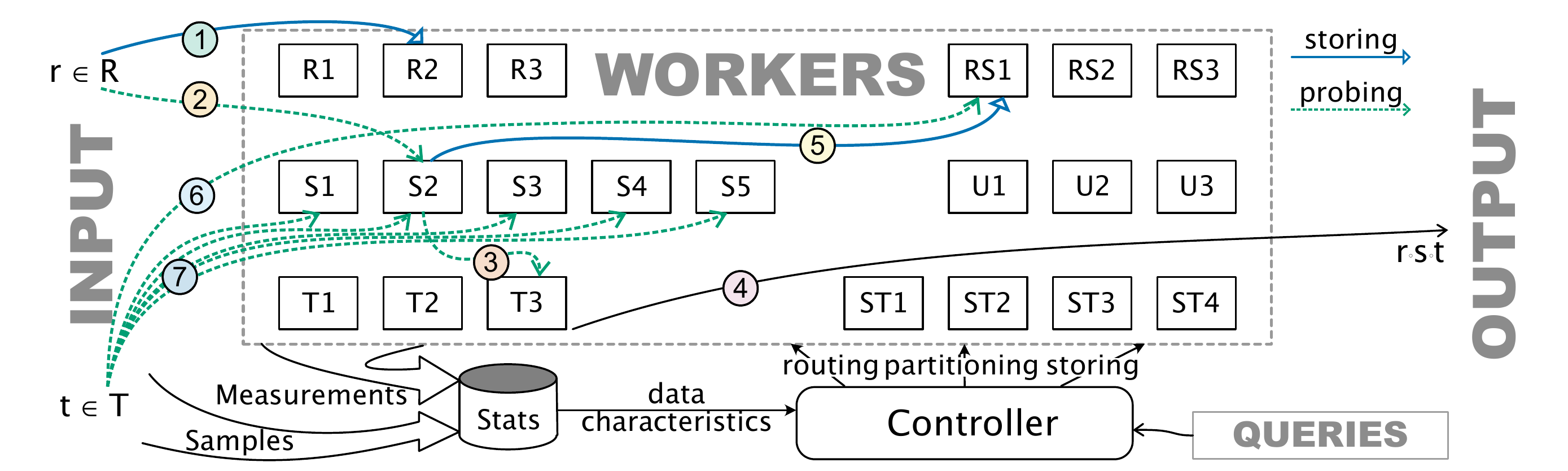}
  \caption{Multi-way join architecture with partitioned relation stores ($R1$, $S2$, etc.), illustrated probing for iterative join result computation, and statistics computation for adaptive optimization of the processing strategy.}\label{fig:big_picture}
\end{figure*}

\section{Distributing Multi-way Stream Joins}\label{sec:distributing_stream_joins}
%

Figure~\ref{fig:big_picture} demonstrates the architecture of our system, \sysname, that we are using for answering multi-way join queries, in this case configured for answering query $R(a), S(a,b), T(b)$.
The input section on the left demonstrates connections to other systems where data items originate, e.g., Kafka-Topics or JMS (Java Message Service) queues.
The center top section contains the workers that perform the join computation.
The graph formed by the workers and their inter connections (not all shown in the figure) is also called a {\it topology}.
Each worker stores a partition of a relation, e.g., $R3$ stores a fraction of the tuples of relation $R$.
We call the joint set of workers for one specific relation a {\bf store}; e.g., $T1$, $T2$, and $T3$ form the $T$-store.
Stores need to be partitioned when the number of tuples to be stored exceeds the local memory limit of a single worker.
For non-equi joins, partitioning can also enable parallel computation of the predicates and thus increase throughput.
However, as we consider only equi joins in this paper, we assume efficient local computation as a given.

Input tuples can arrive at any point in time, like the tuple $r \in R$ shown in the illustration.
It is sent to the partition $R2$ \circled[green]{1} due to its $a$-value where it will be stored and is ready for joining with later arriving tuples.
At the same time, $r$ is sent to partition $S1$, again due to its $a$-value \circled[orange]{2}, where now the join between $r$ and all previously arrived tuples of $S$
is computed (i.e., $\{s | s \in S, s.\tau < r.\tau\})$.
If there are any results, they contain a value for attribute $y$, which can be used for determining the correct partition of relation $T$.
In our example, the intermediate tuple is sent to $T3$ \circled[red]{3}.
At $T3$, the final join result for tuple $r$ is computed and sent to the output \circled[purple]{4}.
In this case, $S$ was partitioned according to attribute $a$, hence, tuples of relation $T$ that do not have a value for $a$ have to be broadcast to all instances of $S$.
This poses the question, which attributes should be used for partitioning.
We address this further below.
Additionally, there is the possibility of materializing intermediate results.
Introducing a $R \Join S$-store to keep intermediate join results can be extremely beneficial in case the intermediate result is very small or the work for computing that intermediate result is very high.
Now, the resulting tuple of $r$ joined with $S$ also needs to be sent to the $R \Join S$-store \circled[yellow]{5}.
This enables tuples from $T$ also to be sent to this store \circled[lightblue]{6} for probing and producing the join result in a single step as opposed to iteratively sending it to the $R$- and $S$-store.

Per se, there is no fixed order in which tuples are sent through the stores of other relations.
For equi joins like in the above example, it seems obvious to send a tuple of relation $R$ first to the $S$-store if $S$ is partitioned by attribute $a$ and not sending it to the $T$ store first.
But even for arbitrary theta joins, the order in which tuples are sent to to-be-joined relations has shown to impact performance, because of vast differences in join selectivities between pairs of relations~\cite{DBLP:conf/bigdataconf/Dossinger019}.
This order is called {\bf probe order}, written as $\langle R, S ,T \rangle$, for example.
It dictates how the tuple of the first mentioned relation, here $R$, is sent to other stores for incrementally computing the join result.
Here, tuples of $R$ are sent first to the $S$-store, then the intermediate result $r \Join S$ is forwarded to the $T$-store.

In general, for a join query $Q$ over $n$ relations $S_1, \dots, S_n$, there is a probe order $\sigma_i$ for each starting relation $S_i$.
This probe order is a permutation over a subset of the installed stores, hence the symbol $\sigma$.
The probe cost for this query is the sum of the cost of executing all individual probe orders:

\begin{equation}\label{eq:pcost}
  \text{PCost}(Q) =
  \sum_{1 \leq i \leq n}
  \sum_{1 \leq j \leq n-1}
  |\bigjoin_{k=1}^{j-1} S_{\sigma_i(k)}| \cdot \frac 1 j \cdot \chi_{\sigma_i(j+1)}.
\end{equation}

Here, $i$ references the input of the current probe order and $j$ the step in that probe order.
In each step, a fraction of the join between the previous relations is produced as intermediate result.
This fraction is $\sfrac{1}{j}$, as in each step the arriving tuple is only joined with tuples that arrived earlier.
The last factor $\chi$ is either $1$ if the partitioning attribute for the $\sigma_i(j)$-store is known, or it is equal to the parallelism of this store since tuples need to be broadcast to every task.
This is illustrated by \circled[blue]{7} in Figure~\ref{fig:big_picture} where $S$ is partitioned according to an attribute unknown to $T$, and, thus, potential join partners can be in every $S$-partition.
In order to compute the correct result, $t$ has to be broadcast to all five partitions of $S$.

\subsection{Simultaneously Answering Multiple Queries}


The optimization of a single multi-way join query like $q = R(a), S(a,b), T(b)$ can already be interpreted as a multi-query optimization problem.
For this, we decompose the query into subqueries $q = q_R \circ q_S \circ q_T$ with $q_R = \{ r \circ s \circ t | r \in R, s \in S, t \in T, r.a = s.a, s.b = t.b, r.\tau > s.\tau, r.\tau > t.\tau\}$ and analogously for $q_S$ and $q_R$.
These subqueries compute only that part of the join result where the tuple of the relevant relation arrives latest and is probed against the previously stored tuples, so they are the result of executing a single probe order.
In general, each subquery can have a different optimal partitioning attribute for the involved stores.
However, each store is only partitioned according to one attribute, and thus we need to take into account the overall probe cost for all subqueries simultaneously.

If multiple join queries are answered and they share inputs, the choice of a partitioning scheme might impact the cost of the probe orders of all queries.
However, there is also a potential for exploiting common parts of probe orders.
For example, consider queries $q_1 = R(a), S(a,b), T(b)$ and $q_2 = R(a), S(a,c), U(c)$.
The intermediate result generated in Figure~\ref{fig:big_picture} from probing $r$ at $S2$ can be used for answering $q_1$ by sending it, as before, to $T3$ (\circled[red]{3}), and also sending it to $U2$ for answering $q_2$ (\patterncircled[orange]{8}).


\section{Optimization using Integer Linear Programming}\label{sec:join_optimization}

An integer linear program (ILP) is in general an optimization problem that determines assignments for a set of variables such that a cost term is minimized (or maximized)~\cite{DBLP:books/daglib/0090562}.
The cost term is the inner product of the user-defined cost for each variable, $c_i$ and the integer variable $x_i$. So, $c_1 x_1 + c_2 x_2 + \dots$ is subject to minimization (or maximization).
Further, these variables have to fulfill a set of constraints, all given in the form of $a_{1,1} x_1 + a_{1,2} x_2 + \dots \geq b_1$.

We follow the approach of~\cite{DBLP:conf/iscis/DokerogluBC14} for formulating a multi-query optimization problem as ILP.
Consider a query $q_i$ for which we have alternative join plans $p_{i,1}, \dots, p_{i,k}$ to choose from.
For each such query $q_i$, we generate equations for the ILP:
\begin{equation}
    \label{eq:eachplanoneoption}
    x_{i,1} + x_{i,2} + \cdots + x_{i,k} = 1
\end{equation}
where $x_{i,j} = 1$ iff plan $j$ is chosen for query $i$.
As the variables  $x_{i,j}$ are integers, these equations are satisfied iff exactly one plan is chosen for each query.
Each plan is composed of multiple tasks which represent the computation of a subresult and have cost assigned.
For example, plan $p_{i,1}$ is composed of tasks $t_1, \dots, t_r$ with cost $c_1, \dots, c_r$ respectively.
Then, we also add equations
\begin{equation}
    \label{eq:costperplan}
    - C x_{i,1} + c_1 x_{t_1} + \cdots + c_r x_{t_r} \geq 0
\end{equation}
where $C := \sum_{i=1\ldots r} c_r$.
Thus, if plan $p_{i,j}$ is chosen, $x_{i,j}$ is set to $1$ and negative cost have to be balanced by selecting all the associated tasks.
The same tasks might appear in candidate plans for different queries, and thus, if such plans are selected, computation can be shared between these plans.
In total, the sum of costs times tasks is subject to minimization.

For our scenario, we need to generate candidate probe orders to choose from, and then translate these choices into topologies.
In order to translate the given query set into an ILP, we first create for each query {\it materializable intermediate results} (MIR) and, based on that, a set of candidate probe orders.
An MIR consists of a subset of the queried relations and the join predicates defined on them such that cross products are avoided.
For example, for query $R(a), S(a,b), T(b)$ the materializable intermediate results would be $(R(a), S(a,b))$ and $(S(a,b), T(b))$ but not $(R(a), T(b))$.

The candidate probe orders are determined using Algorithm~\ref{alg:candidate_probe_order}.
For each relation in the query the recursive subfunction {\tt construct\_rec} is called in order to construct probe orders from head to tail.
It returns all probe orders that can be used to answer $q$ if the starting tuple is the result of joining {\tt head}.
In this subfunction in Line~3, we iterate over all MIRs which are, according to the given query, joinable with the current head.
This way, we avoid producing cross products.
If joining head and $r$ yields a complete result, i.e., all input relations of $q$ are covered, the probe order is completed.
Otherwise, the same function is recursively called to yield all probe orders that start with the previous head joint with $r$.
We assume here that there are no queries which include a cross product.
For this case one can revert back to constructing probe orders as described in~\cite{DBLP:conf/sigmod/DossingerMR19} by adding artificial {\tt true}-join predicates.

\begin{algorithm}[tb]
    \caption{Candidate probe order construction algorithm.}\label{alg:candidate_probe_order}
    \begin{tabbing}
        xxxxxxx\=xxx\=\kill
        {\bf input: } \> query $q$, MIR\\
        {\bf output: } \> candidate probe orders\\
        xxx\=xx\=xx\=xxx\=xxx\= \kill
        {\color{gray} 1} \> {\bf fun} construct\_rec(head)\\
        {\color{gray} 2} \>\> result $\gets []$\\
        {\color{gray} 3} \>\> {\bf for} $r \in $ joinable($q$, head, $\se{MIR}$)\\
        {\color{gray} 4} \>\>\> newHead $\gets$ head $+\ r$ \\
        {\color{gray} 5} \>\>\> {\bf if} newHead is complete \\
        {\color{gray} 6} \>\>\>\> result $\gets$ result $+$ [newHead]\\
        {\color{gray} 7} \>\>\> {\bf else}\\
        {\color{gray} 8} \>\>\>\> result $\gets$ result $+$ construct\_rec[newHead] \\
        {\color{gray} 9} \> \\
        {\color{gray} 10}\> {\bf for} relation {\bf in} query\\
        {\color{gray} 11}\>\> construct\_rec(relation)
    \end{tabbing}
\end{algorithm}

We further need candidate attributes for partitioning of the MIR stores.
For an $r \in \se{MIR}$ these are all attributes which define a join with another relation that is not part of $r$.
To give an example, if for query $q = R(a), S(a,b), T(b)$ the intermediate result $(R(a), S(a,b))$ is materialized, $a$ is {\it not\/} a candidate for partitioning, because there is no join with $T$ that uses this attribute.
However, attribute $b$ is a candidate for partitioning.
This makes sense because all tuples that are sent to the $RS$-store know the value of the $b$ attribute and, hence, can be routed correctly.
Also, partitioning according to $a$ implies that tuples from $T$ need to be broadcast to all $T$-tasks, while with a partitioning according to $b$ this full broadcast is avoided.

While the input relations are always materialized, this is not necessarily the case for intermediate relations.
They are only required if a probe order is selected which also uses those relations.
Thus, for intermediate relations, we also generate probe orders using the subquery for the intermediate result as input to the candidate probe order construction.

\begin{algorithm}[tb]
    \caption{ILP construction procedure.}\label{alg:construction}
    \begin{tabbing}
        xxxxxxx\=xxx\=\kill
        {\bf input: } \> queries $Q$, probe order candidates $C$,\\
        \> partitioning candidates $P$\\
        {\bf output: }\> ilp constraints $A$, optimization goal $G$\\
        xx\=xx\=xx\=xxx\=xxx\= \kill
        {\color{gray} 1} \> $A \gets \{\}$\\
        {\color{gray} 2} \> {\bf for} $q \in Q$, $S \in \mathcal{S}(q)$\\
        {\color{gray} 3} \>\> $p \gets \se{apply\_partitioning}(C[S], P)$\\
        {\color{gray} 4} \>\> {\bf for} $\sigma \in p$\\
        {\color{gray} 5} \>\>\> $A \gets A \cup \se{cost\_constraint}(\sigma)$\\
        {\color{gray} 6} \>\> $A \gets A \cup \se{probe\_order\_constraint}(p)$\\
        {\color{gray} 7} \> $G \gets goal(A)$
    \end{tabbing}
\end{algorithm}
Based on probe order and partitioning candidates, we construct the ILP as shown in Algorithm~\ref{alg:construction}.
In variable $A$, we collect all constraints that the ILP must satisfy.
In Line~2, we iterate over all combinations of queries and possible starting relations to add constraints that select a probe order for each starting relation.
Therefore, the partitioning is applied to the probe order candidates for the starting relation, $C[r]$.
In Line~3, variable $p$ contains probe orders where all MIRs are decorated with the partitioning attribute.
This is necessary for building the cost constraint (Line~5).
The cost values are set according to Equation~\ref{eq:pcost} and in order to compute $\chi$, it is necessary to distinguish between differently partitioned stores.

For computing $\se{probe\_order\_constraint}(p)$, with probe orders $\sigma_1, \dots, \sigma_n \in p$, for each probe order $\sigma_i$ a new variable $x_i \in \{0,1\}$ is introduced.
\begin{align*}
    x_1 + x_2 + \cdots + x_n = 1 \\
\end{align*}
This line resembles Equation~\ref{eq:eachplanoneoption}.
If the probe order identified by $x_i$ contains a materialized intermediate result over relations $1,\dots, l$, this also has to be computed.
Hence for each of the inputs, a probe order which creates the intermediate result needs to be installed, which is made sure by the following constraints:

\begin{align*}
     & - k_1 \cdot x_i + x^\prime_{1,1} + x^\prime_{1,2} + \cdots + x^\prime_{1,k_1}\geq 0  \\
     & \dots                                                                                \\
     & - k_l \cdot x_i + x^\prime_{l,1} + x^\prime_{l,2} + \cdots + x^\prime_{l,k_l} \geq 0 \\
\end{align*}
Here, $k_j$ is set to the number of probe orders required for computing the result starting from relation $j \in 1, \dots, l$.
Variables $x^\prime$ indicate if the probe order for that subquery will be executed.
Since each line needs to be non-negative, it is guaranteed that the intermediate result is actually computed.

The $\se{cost\_constraint(\sigma)}$, which we will model using Equation~\ref{eq:costperplan}, is composed of the cost of all the steps of that probe order.
With $\sigma = \langle S_1, S_2, \dots, S_m \rangle$ these steps are $\rho_1 = S_1 \Join S_2$, $\rho_2 = (S_1 \Join S_2) \Join S_3$ until $(S_1 \dots S_{m-1}) \Join S_m$, so the computation of the partial join result.
Note that computing a step is equivalent to completing a probe order, hence we can also identify steps with probe-order prefixes.
For example, a probe-order prefix $\langle S_1, S_2, S_3 \rangle$ yields the same result as sending the partial result $S_1 \Join S_2$ to $S_3$.
For each step we introduce a {\bf step variable} $y_i$, and it is crucial, that all equal steps used in candidates of other queries get the same variable $y_i$ assigned.
For each of these steps, we also introduce the {\bf step cost} which is innermost term of the inner sum in Equation~\ref{eq:pcost}.
Thus, for each probe order $\sigma$ the following constraint is added:
\begin{align*}
    -PCost(\sigma) \cdot x_1 & + StepCost(\rho_1) \cdot y_1        \\
    + \cdots                 & + StepCost(\rho_m) \cdot y_m \geq 0
\end{align*}

In Line~7, the goal is set.
The goal is derived from the step cost and step variables of the previously added constraints:
\begin{align*}
    \min \sum_{i = 1 \dots m} StepCost(\rho_i) \cdot y_i
\end{align*}
As $y_i \in \{0,1\}$, the value of the sum is only affected by the variables set to 1.
Only combinations of variables $y_i$ can be set to $1$ such that all queries have all necessary probe orders for computing their results.
Thus, a solution that minimizes this term can also be translated to a correctly working topology.
This topology needs to be deployed to a stream processor like Apache Storm where it processes the query.

\subsubsection{ILP Creation Example}
Consider queries $q_1 = R(b), S(b,c), T(c)$ and $q_2 = S(c), T(c,d), U(d)$.
In Figure~\ref{fig:walkthrough}, we see first the materializable intermediate results composed of the input relations as well the intermediate results.
E.g., $RS$ stands here for the result of the subquery $q_{RS} = R(a,b),S(b,c)$.
Since we have potential intermediate results, they also need to be created, and thus, probe orders for them have to be installed as well.
The probe orders are listed next.
There is one probe order per input relation of each query.
For example, $q_1$ consists of three inputs and hence, three sets of probe orders are created and one of the candidates of each set need to be used.

Thereafter, the partitioning is applied to the probe orders.
In Figure~\ref{fig:walkthrough}, we only show the options for probe orders for $q_1$ and $R$.
Here it is interesting to see, that also partitioning which is not beneficial to the current query is included.
For example, the probe order $\langle R, S[b], T[d] \rangle$ indicates that the $S$-store is partitioned according to attribute $b$ and the $T$-store is partitioned according to $d$.
If this probe order is installed, a tuple from $R$ after it probed the $S$-store needs to be broadcast to all $T$-workers in order to compute the result for $q_1$, because this tuple does not contain the value of attribute $d$.
The partitioning of $T$ according to $d$ is only useful for $q_2$.

Finally, the constraints for the ILP are added.
The first constraint requires that exactly one from the probe order candidates $\sigma_1$ to $\sigma_6$ is chosen.
For this we add an ILP variable $x_i$ for each $\sigma_i$ that takes values in $\{0,1\}$.
Then, we need to make sure, that for probe orders which include intermediate results, these intermediate results are actually computed.
The next constraint showcases this for $\sigma_5$.
In this probe order, $R$-tuples are sent to the $ST$-store which is partitioned according to $b$ for probing.
To do so, the $ST$-store needs to be installed and also kept up to date with this intermediate result.
In turn, probe orders for computing $S \Join T$ need to be installed.
In this case, there are four probe orders, one for sending $S$ to the $T$-store and one for sending $T$ to the $S$-store and each store can be partitioned according to two attributes.
Out of these probe orders we need two (one for each relation) and thus we add constraints 2 and 3.
Actually, the computation of the intermediate result is independent from the partitioning of this result's store.
Thus, the same intermediate result computation can be used for $\sigma_6$.
We then need to make sure that each probe order, if it is chosen, is computed correctly.
Probe order $\sigma_1$, for example, has the prefix $\sigma_7$.
In this example, the probe order steps and prefixes are used interchangeably.
Now, we add constraint~4 where ILP variables for each step in that probe order are set, $y_7$ and $y_1$.
These variables are associated with the step cost for $\sigma_7$, i.e., sending tuples from $R$ to the $S$-store which is partitioned by $b$, and the step cost for $\sigma_1$, i.e., sending tuples from the $S$ store to the $T$ store which is partitioned by $c$.
In constraint 5, the next probe order has the same first step, and thus, it is crucial that the same variable $y_7$ is put into the ILP.
The optimization goal of the ILP is then to minimize the sum of the step costs of all used steps.

\begin{figure}[tb]
    $q_1 = R(b), S(b,c), T(c)$, $q_2 = S(c), T(c,d), U(d)$

    $\se{MIR} = R, S, T, U, \se{RS}, \se{ST}, \se{TU}$

    {\bf Candidate probe orders:}

    \begin{minipage}{.49\columnwidth}
        for $q_1$:
        \begin{itemize}
            \item {\bf R}: $\langle R, S, T \rangle$, $\langle R, \se{ST} \rangle$
            \item {\bf S}: $\langle S, T, R \rangle$, $\langle S, R, T \rangle$
            \item {\bf T}: $\langle T, S, R \rangle$, $\langle T, \se{RS} \rangle$
        \end{itemize}
    \end{minipage}
    \begin{minipage}{.49\columnwidth}
        for $q_2$:
        \begin{itemize}
            \item {\bf S}: $\langle S, T, U \rangle$, $\langle S, \se{TU} \rangle$
            \item {\bf T}: $\langle T, U, S \rangle$, $\langle T, S, U \rangle$
            \item {\bf U}: $\langle U, T, S \rangle$, $\langle U, \se{ST} \rangle$
        \end{itemize}
    \end{minipage}
    \begin{minipage}{.49\columnwidth}
        for $q_{RS}$:
        \begin{itemize}
            \item {\bf R}: $\langle R, S \rangle$
            \item {\bf S}: $\langle S, R \rangle$
        \end{itemize}
    \end{minipage}
    \begin{minipage}{.49\columnwidth}
        for $q_{ST}$:
        \begin{itemize}
            \item {\bf S}: $\langle S, T \rangle$
            \item {\bf T}: $\langle T, S \rangle$
        \end{itemize}
    \end{minipage}
    \begin{minipage}{.49\columnwidth}
        for $q_{TU}$:
        \begin{itemize}
            \item {\bf T}: $\langle T, U \rangle$
            \item {\bf U}: $\langle U, T \rangle$
        \end{itemize}
    \end{minipage}
    \vspace{3pt}

    Probe orders with partitioning for $q_1$ and $R$, including probe order prefixes:\\

    \begin{minipage}{.315\columnwidth}
        \begin{itemize}
            \item[$\sigma_1$] $\langle R, S[b], T[c] \rangle$
            \item[$\sigma_2$] $\langle R, S[c], T[c] \rangle$
            \item[$\sigma_3$] $\langle R, S[b], T[d] \rangle$
        \end{itemize}
    \end{minipage}
    \hfill
    \begin{minipage}{.315\columnwidth}
        \begin{itemize}
            \item[$\sigma_4$] $\langle R, S[c], T[d] \rangle$
            \item[$\sigma_5$] $\langle R, \se{ST[b]} \rangle$
            \item[$\sigma_6$] $\langle R, \se{ST[d]} \rangle$
        \end{itemize}
    \end{minipage}
    \hfill
    \begin{minipage}{.3\columnwidth}
        \begin{itemize}
            \item[$\sigma_7$] $\langle R, S[b]\rangle$
            \item[$\sigma_8$] $\langle R, S[c]\rangle$
        \end{itemize}
        \vfill
    \end{minipage}

    \vspace{3pt}
    {\bf Constraints:}
    \begin{enumerate}
        \item $x_{\sigma_1} + x_{\sigma_2} + x_{\sigma_3} + x_{\sigma_4} + x_{\sigma_5} + x_{\sigma_6} = 1$ (one probe order)
        \item $-2 x_{\sigma_5} + x_{\sigma^\prime_1} + x_{\sigma^\prime_2} \geq 0$
        \item $-2 x_{\sigma_5} + x_{\sigma^\prime_3} + x_{\sigma^\prime_4} \geq 0$ (subqueries for $\sigma_9$)
        \item $-PCost(\sigma_1) \cdot x_{\sigma_1} + StepCost(\sigma_7) \cdot y_7$\\ $ + StepCost(\sigma_1) \cdot y_1 \geq 0$
        \item $-PCost(\sigma_3) \cdot x_{\sigma_3} + StepCost(\sigma_7) \cdot y_7$\\ $ + StepCost(\sigma_3) \cdot y_3 \geq 0$
        \item[] $\dots$
    \end{enumerate}

    {\bf Optimization goal:}
    \begin{align*}
        \min\quad & StepCost(\sigma_1) y_{\sigma_1} + StepCost(\sigma_3) y_{\sigma_3} \\
                  & +  StepCost(\sigma_7) y_{\sigma_7} + \dots
    \end{align*}

    \caption{Deriving ILP for queries $q_1$ and $q_2$.}\label{fig:walkthrough}
\end{figure}

\subsubsection{Multi Query Optimization Example}
In this example, we only focus on choosing probe orders and ignore materializing subqueries and partitioning.
Thus, we ignore additional cost for broadcasting and do not write the partitioning attribute.
Consider the queries $q_1 = R(a), S(a,b), T(b)$ and $q_2 = S(b), T(b,c), U(c)$ where each relation streams at a rate of $100$ tuples per time unit and the join between $S$ and $T$ produces $150$ intermediate results, while the other join produces only $100$ intermediate results.
We now focus on what happens with relations $S$ in $q_1$ and $T$ in $q_2$.
Optimizing each query individually, we would install the probe orders $\langle S, R, T\rangle$ and $\langle T, U, S\rangle$ in order to avoid the more expensive intermediate join between $S$ and $T$, and send in total $475$ tuples for probing tuples in each query, thus $950$ tuples in total.
Since for answering $q_1$ (respectively, $q_2$) correctly tuples must to be sent from $T$ to $S$ ($S$ to $T$), we can exploit this and instead install probe orders $\langle T, S, U\rangle$ ($\langle S, T, R\rangle$) for $q_2$ ($q_1$).

For the optimization problem we assign variables for the steps in the probe order, e.g., $x_{RS}$ for the cost of sending $R$-tuples to the $S$-store for probing, or $X_{RST}$ for the cost of sending the intermediate result of $R \Join S$ to the $T$-store for probing. The cost associated with these variables is $100$ for all first steps, and $75$ for joins between $S$ and $T$ and $50$ for the other joins (cf. Formula~\ref{eq:pcost}). For $q_1$ and starting relation $S$ the following constraint rows are added to the ILP:

\begin{align*}
    x_1 + x_2                          & = 1    \\
    -150 x_1 + 100 x_{SR} + 50 x_{SRT} & \geq 0 \\
    -175 x_2 + 100 x_{ST} + 75 x_{STR} & \geq 0
\end{align*}

$x_1$ stands for the probe order $\langle S, R, T \rangle$ and $x_2$ for the probe order $\langle S, T, R \rangle$. The first line makes sure that only one of these variables can be 1 and this variable determines which of the probe orders will be installed in the running topology. The second line enforces that if $x_1$ is set to $1$, then also $x_{SR}$ and $x_{SRT}$ are set to $1$.

For $q_2$ and starting relation $S$ there is only one probe order, thus the following constraint rows are added to the ILP:

\begin{align*}
    x_3                                & = 1    \\
    -150 x_3 + 100 x_{ST} + 75 x_{STU} & \geq 0 \\
\end{align*}

Essentially, this leaves no choice: $x_3$ has to be set, and consequently also $x_{ST}$ and $x_{STU}$ have to be set, and thus $S$ tuples need to be sent to $T$ and afterwards to $U$ in order to produce all desired join results. The optimization goal then includes the here mentioned cost-variables and more which are not shown for clarity:

\begin{align*}
    \min\quad 100 x_{SR} + 50 x_{SRT} + 100 x_{ST} + 75 x_{STR} + 75 x_{STU}
\end{align*}

As discussed, $x_{ST}$ and $x_{STU}$ need to be set due to $q_2$. This way, selecting the probe order $x_2$ (and thus setting $x_{STR}$ to 1) adds only $75$ to the cost. Selecting $x_1$, on the other hand, requires $x_{SR} = x_{SRT} = 1$ and adds $150$ to the cost. Hence, the locally---for query $q_1$ in isolation---suboptimal probe order $x_2$ is chosen and an overall lower number of tuples need to be sent around.

\subsection{Analysis}
The number of materializable intermediate results of a query over $n$ relations is in the worst case $2^n$ when the query graph is a clique, i.e., for every pair of relations there is a join predicate. For example for a linear query, the size of MIRs is the number of consecutive subsequences of a word of length $n$, so only $n(n+1)$. The number of candidate probe orders per query and relation is, in the worst case, the number of permutations of these subsequences times the number of partitioning options.
This all heavily depends on the query. For example, for a linear query there are $2^{n-2}$ and a star query has only $n-1$ partitions to choose from.
The number of ILP variables is then for all queries the sum of the amount of candidates for each query, as well as the prefixes of the probe orders.

\subsection{Transformation to Executable Strategies.}

The result of the ILP optimization is the assignment of probe order variables (and step variables, but we can ignore them). We now detail on how to construct a topology of compute tasks for actually computing the query.

\begin{figure}[tb]
    \centering
    \begin{tikzpicture}[xscale=0.5]



        \node (po1) {$\sigma_1: \langle R, S[d], T[b] \rangle$};
        \node[below=0 of po1] (po2) {$\sigma_2: \langle R, S[d], W[e] \rangle$};
        \node[below=0 of po2] (po3) {$\sigma_3: \langle R, U[a] \rangle$};

        \node[right=of po2] (R4) {R};
        \node[above right=0 and 1 of R4] (S4) {S[d]} edge[<-] node[above] {$s_1$} (R4);
        \node[below right=0 and 1 of R4] (U4) {U[a]} edge[<-] node[below] {$s_2$} (R4);
        \node[above right=0 and 1 of S4] (T4) {T[b]} edge[<-] node[above] {$s_3$} (S4);
        \node[below right=0 and 1 of S4] (W4) {W[e]} edge[<-] node[below] {$s_4$} (S4);
    \end{tikzpicture}
    \caption{Three probe orders for the same starting relation merged into a probe tree.}\label{fig:probetree}
\end{figure}
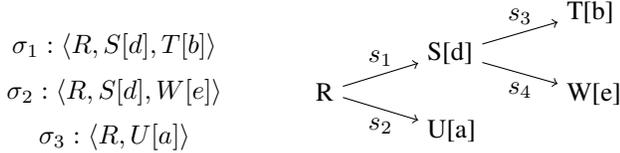

The probe orders with variables assigned $1$ are the probe orders that should be used in the actual query execution. We merge probe orders into probe trees, as illustrated in Figure~\ref{fig:probetree}. Here, we see several probe orders for the starting relation $R$. Since $\sigma_1$ and $\sigma_2$ both have the same first step, probing the $S$-store, they are represented by the edge from $R$ to the node with label $S[d]$. Multiple outgoing edges in this graph indicate that a tuple is copied and sent to both targeted stores. This is done for all probe orders, such that we end up with a forest of such probe trees. For each distinct label of the inner nodes, a store is introduced in the topology. This way, nodes with the same label in different probe trees refer to the same store and data is not stored redundantly. For the roots, ingestion methods (in case of Storm these are Spouts) need to be installed. For each edge of a probe graph, a new, unique, edge label is introduced.

With help of these edge labels, rules are registered at all stores.
These rules define the behavior of the store for a received tuple based on the incoming edge label.
The sending store is not enough, as there might be tuples from different probe trees sent from one to the other store.
These tuples stem from different (sub)relations and the probe result is sent to different stores for further processing, so we use the edge labels instead.
A rule follows the pattern {\it if tuple arrives from edge $E_{in}$, probe using predicate $P$, and send result (if any) to $E_{out}$}.
All rules registered to a store are organized in a ruleset.
On each arriving tuple this ruleset is consulted for deciding how to proceed with the tuple.
During runtime, Algorithm~\ref{alg:handle_nonadaptive} is used to decide on a worker how to process a tuple: in Line~2, the matching rules for the incoming edge are extracted.
Since this is done for every tuple, this must happen quickly, so the ruleset is organized as hash map keyed by the incoming edge labels.
Then the type of the rule decides how the arriving tuple has to be handled.
If the rule is a {\it store rule}, like in Line~4, the arriving tuple is added to the local store of arrived tuples, and is ready for other later arriving probe tuples.
These arrive over edges where a {\it probe rule} is registered.
If such a tuple arrives, Line~5 makes sure it probes with the previously arrived tuples of the stored relation.

A probe rule contains a description of the way of accessing the tuples.
For example, a tuple sent via $s_3$ in Figure~\ref{fig:probetree} contains a partial result of $R \Join S$, and the $T$-store contains the previously arrived tuples of $T$.
For the local probe handling at workers it is irrelevant how the store is partitioned.
Consider here that the probe should determine join partners for the predicate $R.b = T.c$.
The probe rule accesses the $R.b$-attribute of the incoming tuple and needs to find all stored tuples with the same value in $T.c$ for creating join results.
For each distinct attribute access in a store, indices are created locally for efficiently answering probe request.

\begin{algorithm}[tb]
    \caption{Non-adaptive version of incoming tuple handling procedure.}\label{alg:handle_nonadaptive}
    \begin{tabbing}
        xxx\=xx\=xx\=xxx\=xxx\= \kill
        {\color{gray} 1} \> {\bf fun} handle($e_{in}$, tuple)\\
        {\color{gray} 2} \>\> rules $\gets$ ruleset[$e_{in}$]\\
        {\color{gray} 3} \>\> {\bf for} rule {\bf in} rules\\
        {\color{gray} 4} \>\>\> {\bf switch} type(rule)\\
        {\color{gray} 4} \>\>\>\> {\bf case} StoreRule: store(tuple)\\
        {\color{gray} 5} \>\>\>\> {\bf case} ProbeRule: probe(tuple)
    \end{tabbing}
\end{algorithm}

\section{Adaptive Join Processing}\label{sec:adaptive_probe_ordering}

As data characteristics or query work loads change, it might be beneficial to switch to a new strategy.
We achieve the goal to perform this switch without downtime or loss of results in the meantime
by dividing time into epochs and making the configuration of all components depending on these epochs.

\subsection{Epoch-Based Configuration}

\begin{figure}[tb]
    \includegraphics[width=\columnwidth]{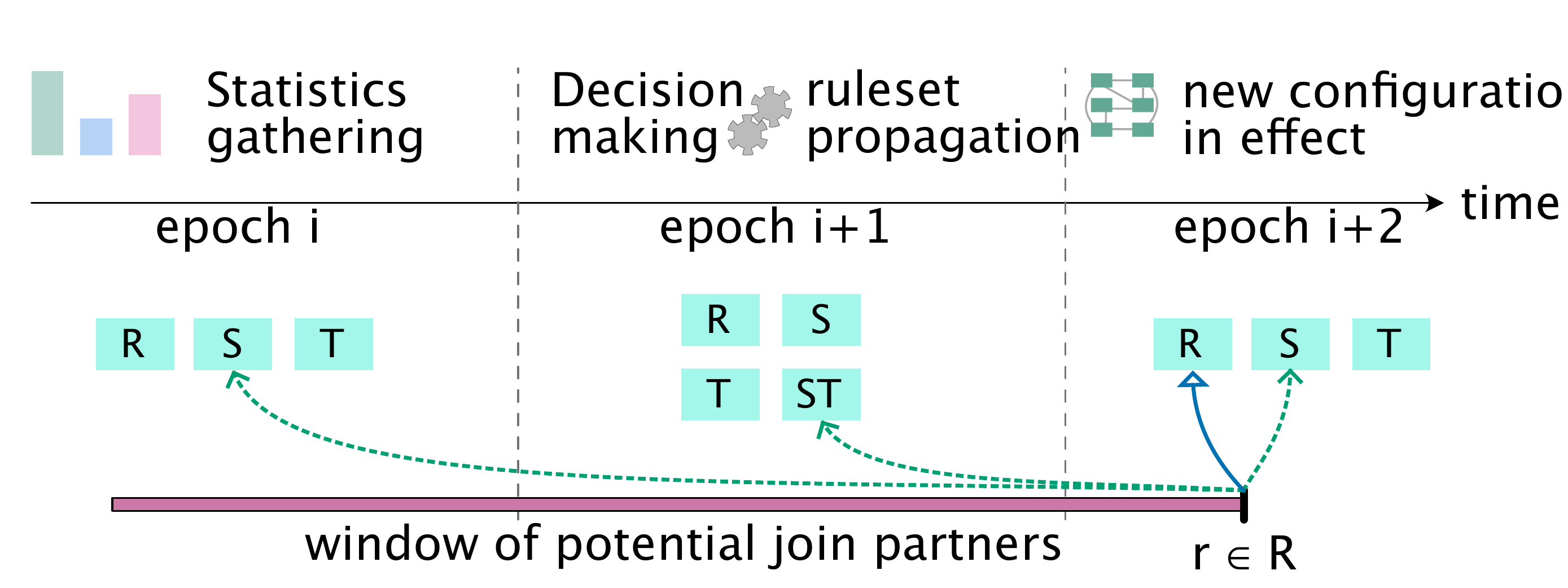}
    \caption{{\it Above:\/} Changes in statistics gathered during one epoch impact the epoch after the next one. {\it Below:\/} tuple $r \in R$ arrives and can find join partners depending on the stores installed in the candidate epochs.}\label{fig:epochs}
\end{figure}

Time is divided into non-overlapping epochs $e_1, e_2, \dots$.
An epoch has a starting timestamp and is considered the current epoch until another epoch with a later timestamp is created.
For each epoch, the data sampled from the epoch is used to create epoch-local data characteristics.
This is done in the next epoch, and so changes can be decided for the epoch after that.
Figure~\ref{fig:epochs} illustrates this: during epoch $i$ sample data is gathered from the inputs.
When epoch $i+1$ starts, the statistics from $i$ are evaluated and fed into the ILP optimizer.
If the optimization result differs from the previous one, a new configuration is created.
This configuration is sent to all workers to be active starting at epoch $i+2$.

In this example, there are two targets for the first probe, and in general there can be more.
Thus, the task receiving the input tuple needs to keep track of where tuples need to be sent.
Algorithm~\ref{alg:handle_input} demonstrates how this is done.
In Line~2, we determine the target epochs where join partners according to the windows in the query can be.
This also depends on the queries installed in the system.
In Figure~\ref{fig:epochs}, for tuple $r$ the target epochs are $i$, $i+1$, and $i+2$, and the receivers are the $S$ and the $ST$-store.
This could be the result of the optimizer deciding for epochs $i$ and $i+2$ to use probe orders $\langle R, S, T \rangle$ and for epoch $i+1$ epoch $\langle R, ST \rangle$.
In Line~3, we iterate over the receivers and then emit the tuple in Line~4 to the receivers and also send the target epoch.
This epoch variable signals the state of the stores the probe tuple wants to see.

This is reflected in the changed handle function, also shown in Algorithm~\ref{alg:handle_input}.
Here each tuple arrives annotated with an epoch.
Using this epoch, we get the ruleset that is valid for this epoch in Line~7.
If there are store or probe rules, we also store or probe with respect to this epoch in Lines~10 and 11.
This means, that also for each epoch, an independent container is created on each worker together with all aforementioned indexes.
If at the end of probing a result is observed, the receivers of the next step depend on the epochs determined by the originating tuple's timestamp.
In the end, the entire result consists of the union of the results of all covered epochs.

As the query's window is not aligned with the epochs, the workers need to check not only the join predicate, but also that the window condition is satisfied.

\begin{algorithm}[tb]
    \caption{Adaptive version of handling procedures for tuples of input relations and intermediate tuples.}\label{alg:handle_input}
    \begin{tabbing}
        xxx\=xx\=xx\=xxx\=xxx\= \kill
        {\color{gray} 1} \> {\bf fun} handle\_input(tuple)\\
        {\color{gray} 2} \>\> {\bf for} epoch {\bf in} get\_epochs\_for(tuple)\\
        {\color{gray} 3} \>\>\> {\bf for} receiver {\bf in} receivers of target\_epochs\\
        {\color{gray} 4} \>\>\>\> emit(receiver, epoch, tuple)\\
        {\color{gray} 5} \\
        {\color{gray} 6} \> {\bf fun} handle($e_{in}$, epoch, tuple)\\
        {\color{gray} 7} \>\> rules $\gets$ ruleset[epoch, $e_{in}$]\\
        {\color{gray} 8} \>\> {\bf for} rule {\bf in} rules\\
        {\color{gray} 9} \>\>\> {\bf switch} type(rule)\\
        {\color{gray} 10} \>\>\>\> {\bf case} StoreRule: store(epoch, tuple)\\
        {\color{gray} 11} \>\>\>\> {\bf case} ProbeRule: probe(tuple)
    \end{tabbing}
\end{algorithm}

\subsection{Supporting Query Changes}

So far, the description focused on a given set of queries and how to adapt to changing data characteristics.
In a long-standing streaming system, users also want to install new queries or remove old ones when they are not interesting anymore, which also captures updating a query.
When a new query is installed, at the next run of the optimization procedure, it is also considered and corresponding probe orders will be generated.
Hence, results can start being reported as soon as the new configuration is installed.

\begin{figure}[tb]
    \includegraphics[width=\columnwidth]{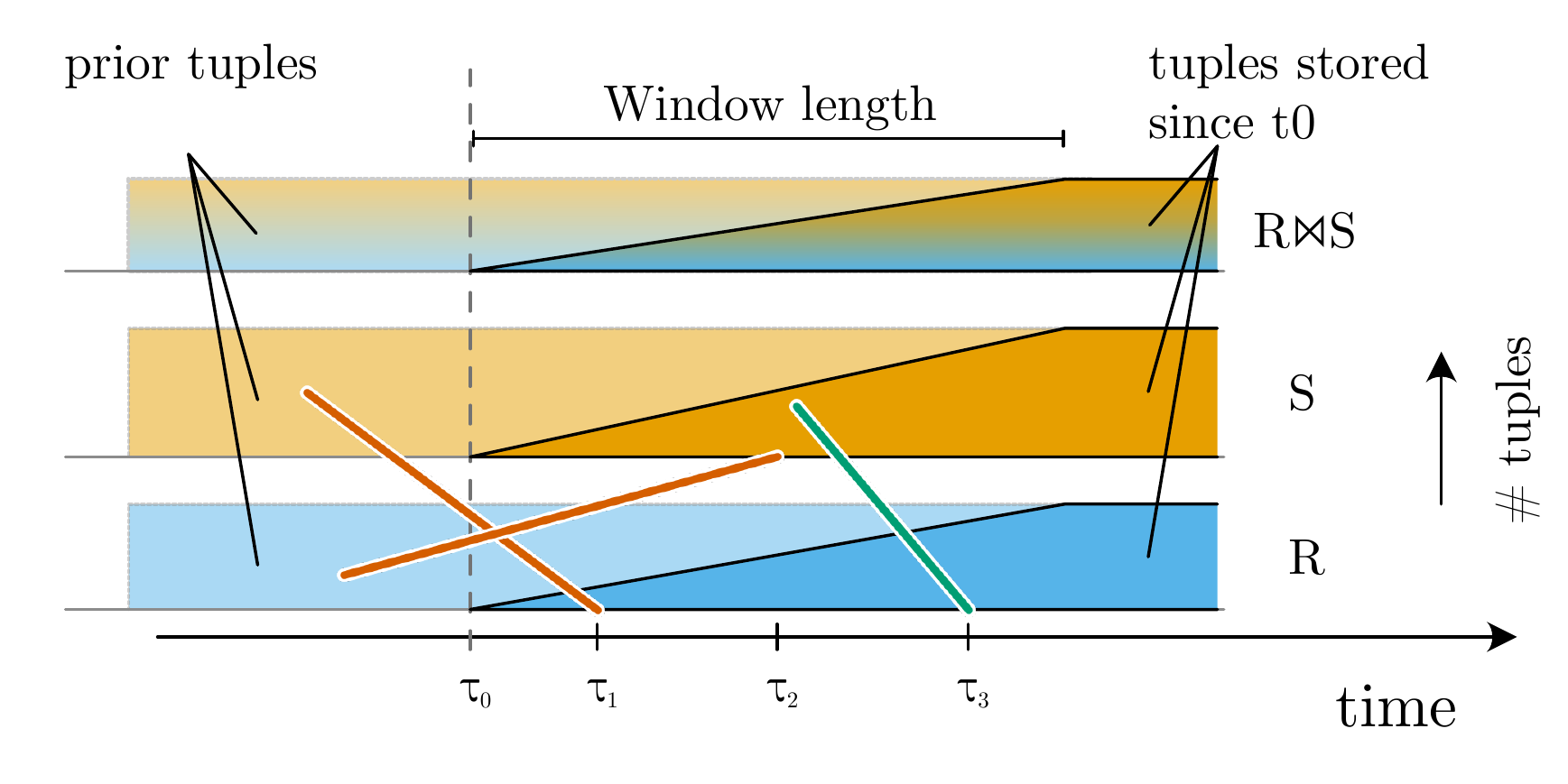}
    \caption{The result of query $R \Join S$ installed at $\tau_0$, when it can only access tuples arrived later than $\tau_0$ or also prior tuples.}\label{fig:newqueries}
\end{figure}

Typically, if a system starts answering a new query, the first window size does not contain all data.
This is because only after the query is installed, tuples are started to be collected in operators for joining.
Consider the scenario in Figure~\ref{fig:newqueries} where at time $\tau_0$ a new query for joining $R \Join S$ is installed.
If streamed relations $R$ and $S$ were only to be observed since $\tau_0$, consequently only these tuples can be probed against.
This means, if at $\tau_1$, a tuple from $R$ arrives, and it is probed against the $S$-store, it is not possible for the system to match these tuples from $S$ that satisfy the join predicate and the window condition, but were observable in the original data stream before $\tau_0$, as indicated by the red line from $\tau_1$ into the past.
Vice versa for the tuple arriving at $\tau_2$ which cannot meet the theoretical join partner from $R$.
If at time $\tau_3$ the tuple arrives and a join partner was arriving after $\tau_0$ in the probed stream, this partial result can be made.
Thus, only after waiting a full window length, such a system can provide complete answers.
If a system is continuously running, and as it is answering other queries, the state used for the other queries is available to a new one.
This means, the registered stores can be exploited to provide complete answer for new queries quicker.

If for all but one inputs of a query, stores are registered, we compute probe orders for all epochs that overlap with the current window, and append these probe orders to the worker's configurations.
This way, we can instantly begin answering all desired join results, and avoid the bootstrap problem of having incomplete statistics.

When a query is not needed anymore, it is removed from the optimizer input.
But that also means, that previous store windows might not be needed anymore.
A reference counting strategy determines the number of queries a store is serving.
As soon as this counter drops to zero, the store is deregistered.




\section{Experiments}\label{sec:experiments}

The experiments are organized  into three sections.
First, we investigate the overall performance of the methods provided in this paper, that is, adaptive multi query optimization.
Second, we specifically look at single-query performance to understand the benefits of adapting query plans to changing data characteristics.
Last, the impact of input sizes to the ILP performance is investigated in detail.

We implemented the described routines as extensions to CLASH~\cite{DBLP:conf/sigmod/DossingerMR19} in Kotlin 1.4, publicly available on GitHub\footnote{\url{https://github.com/clash-streaming/clash}}, while Gurobi 9.0.0rc2\footnote{\url{https://www.gurobi.com/}} is used as the solver for the ILPs.
The optimized query plans are translated by CLASH into Apache Storm v2.2.0~\cite{website:storm} topologies and executed on OpenJDK 11 running on a compute cluster of 8 machines. Each machine has 128GB DDR3 memory and two Intel Xeon CPUs \@1.7 GHz with 6 cores.
This means, we could run up to 96 workers with 10GB memory in parallel.
The cluster nodes are connected using 10Gbs ethernet network.
Input data is consumed from and output is written to Kafka over the same network; the state of the stores is kept in the main memory of the worker processes.

\subsection{Multi-Query Performance}

The following alternatives to processing bulks of queries are considered for comparison:
\begin{enumerate}
  \item Several independent Apache Flink~\cite{website:flink} Jobs, one for each query, are initiated. We refer to this strategy as {\bf Flink Independent} (FI).
  \item Analogously for Apache Storm topologies, coined {\bf Storm Independent} (SI).
  \item A naive multi query optimization strategy where each query is optimized individually with common subplans being executed only once and shared, in Flink, coined {\bf Flink Shared} (FS) and
  \item likewise for Storm, {\bf Storm Shared} (SS).
  \item Lastly, our approach of global optimization: {\bf CLASH-MQO} (CMQO).
\end{enumerate}

We used the well-known TPC-H data set~\cite{website:tphc} with a scale factor of 10.
We create join queries based on present primary, foreign keys and, additionally, type compatible data of TPC-H, which means that two columns can be used for joining if they contain equal values.
This leads to a mixture of common primary-foreign-key style joins, high-selectivity joins (e.g., on `lineitem.linestatus' and `orders.orderstatus' where the domain consists only of {\tt F}, {\tt O}, and {\tt P}), and low-selectivity joins (e.g., on `customer.custkey' and `nation.nationkey' where only customer tuples with the lowest keys find a join partner).
Using these potential joins, we construct queries by selecting a random relation and then randomly adding joins until the desired query size is reached.

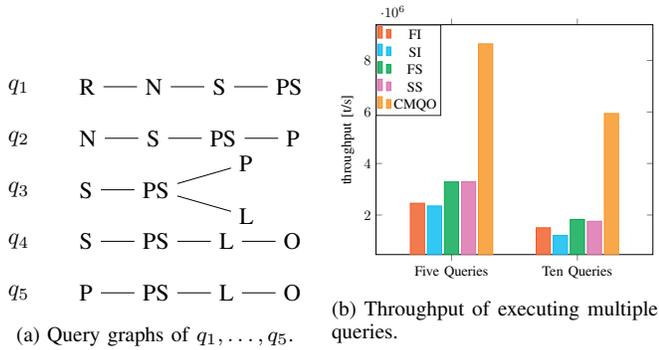
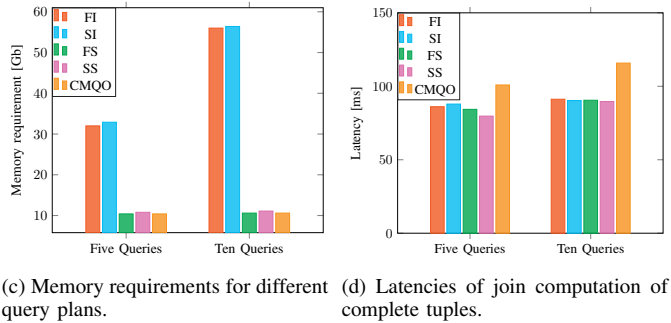
\begin{figure}[tb]
  \centering
  \begin{subfigure}[b]{0.49\columnwidth}
    \resizebox{\columnwidth}{!}{
      \begin{tikzpicture}[node distance=0.3 and 0.5]
        \node(q1) {$q_1$};
        \node[right=of q1] (R1) {R};
        \node[right=of R1] (N1) {N} edge (R1);
        \node[right=of N1] (S1) {S} edge (N1);
        \node[right=of S1] (PS1) {PS} edge (S1);

        \node[below=of q1] (q2) {$q_2$};
        \node[right=of q2] (N2) {N};
        \node[right=of N2] (S2) {S} edge (N2);
        \node[right=of S2] (PS2) {PS} edge (S2);
        \node[right=of PS2] (P2) {P} edge (PS2);

        \node[below=of q2](q3) {$q_3$};
        \node[right=of q3] (S3) {S};
        \node[right=of S3] (PS3) {PS} edge (S3);
        \node[above right=0 and 0.8 of PS3,yshift=-0.1cm] (P3) {P} edge (PS3);
        \node[below right=0 and 0.8 of PS3,yshift=0.1cm] (L3) {L} edge (PS3);

        \node[below=of q3](q4) {$q_4$};
        \node[right=of q4] (S4) {S};
        \node[right=of S4] (PS4) {PS} edge (S4);
        \node[right=of PS4] (L4) {L} edge (PS4);
        \node[right=of L4] (O4) {O} edge (L4);

        \node[below=of q4] (q5) {$q_5$};
        \node[right=of q5] (P5) {P};
        \node[right=of P5] (PS5) {PS} edge (P5);
        \node[right=of PS5] (L5) {L} edge (PS5);
        \node[right=of L5] (O5) {O} edge (L5);
      \end{tikzpicture}
    }
    \caption{Query graphs of $q_1, \dots, q_5$.\\}\label{fig:queries_q5}
  \end{subfigure}
  \hfill
  \begin{subfigure}[b]{0.49\columnwidth}
    \resizebox{\columnwidth}{!}{
      \begin{tikzpicture}
        \begin{axis}[
            ybar,
            enlarge x limits=.6,
            legend style={anchor=north west, at={(0,1)}, inner xsep=1pt, inner ysep=1pt},
            ylabel={throughput [t/s]},
            symbolic x coords={Five Queries,Ten Queries},
            xtick=data
          ]
          \addplot[color=red, fill=red, fill opacity=0.8] coordinates { (Five Queries, 2456723) (Ten Queries, 1503246) };
          \addplot[color=lightblue, fill=lightblue, fill opacity=0.8] coordinates { (Five Queries, 2349567) (Ten Queries, 1207684) };
          \addplot[color=green, fill=green, fill opacity=0.8] coordinates { (Five Queries, 3283738) (Ten Queries, 1826458) };
          \addplot[color=purple, fill=purple, fill opacity=0.8] coordinates { (Five Queries, 3289487) (Ten Queries, 1756405) };
          \addplot[color=orange, fill=orange, fill opacity=0.8] coordinates { (Five Queries, 8642378) (Ten Queries, 5940132) };
          \legend{FI,SI,FS,SS,CMQO}
        \end{axis}
      \end{tikzpicture}
    }
    \caption{Throughput of executing multiple queries.}\label{fig:throughput_q5_10}
  \end{subfigure}
  \hfill

  \begin{subfigure}[b]{0.49\columnwidth}
    \resizebox{\columnwidth}{!}{
      \begin{tikzpicture}
        \begin{axis}[
            ybar,
            enlarge x limits=.6,
            legend style={anchor=north west, at={(0,1)}, inner xsep=1pt, inner ysep=1pt},
            ylabel={Memory requirement [Gb]},
            symbolic x coords={Five Queries,Ten Queries},
            xtick=data
          ]
          \addplot[color=red, fill=red, fill opacity=0.8] coordinates { (Five Queries, 32.0) (Ten Queries, 56.0) };
          \addplot[color=lightblue, fill=lightblue, fill opacity=0.8] coordinates { (Five Queries, 32.9) (Ten Queries, 56.4) };
          \addplot[color=green, fill=green, fill opacity=0.8] coordinates { (Five Queries, 10.4) (Ten Queries, 10.6) };
          \addplot[color=purple, fill=purple, fill opacity=0.8] coordinates { (Five Queries, 10.8) (Ten Queries, 11.1) };
          \addplot[color=orange, fill=orange, fill opacity=0.8] coordinates { (Five Queries, 10.4) (Ten Queries, 10.6) };
          \legend{FI,SI,FS,SS,CMQO}
        \end{axis}
      \end{tikzpicture}
    }
    \caption{Memory requirements for different query plans.}\label{fig:memory_q5_10}
  \end{subfigure}
  \hfill\begin{subfigure}[b]{0.49\columnwidth}
    \resizebox{\columnwidth}{!}{
      \begin{tikzpicture}
        \begin{axis}[
            ybar,
            enlarge x limits=.6,
            legend style={anchor=north west, at={(0,1)}, inner xsep=1pt, inner ysep=1pt},
            ylabel={Latency [ms]},
            symbolic x coords={Five Queries,Ten Queries},
            xtick=data,
            ymin=0,
            ymax=150,
          ]
          \addplot[color=red, fill=red, fill opacity=0.8] coordinates { (Five Queries, 86.1) (Ten Queries, 91.2) };
          \addplot[color=lightblue, fill=lightblue, fill opacity=0.8] coordinates { (Five Queries, 87.9) (Ten Queries, 90.3) };
          \addplot[color=green, fill=green, fill opacity=0.8] coordinates { (Five Queries, 84.3) (Ten Queries, 90.5) };
          \addplot[color=purple, fill=purple, fill opacity=0.8] coordinates { (Five Queries, 79.7) (Ten Queries, 89.6) };
          \addplot[color=orange, fill=orange, fill opacity=0.8] coordinates { (Five Queries, 100.9) (Ten Queries, 115.8) };
          \legend{FI,SI,FS,SS,CMQO}
        \end{axis}
      \end{tikzpicture}
    }
    \caption{Latencies of join computation of complete tuples.}\label{fig:latency_q5_10}
  \end{subfigure}
  \caption{Multi-Query Performance on TPC-H data.}\label{fig:experiments}
\end{figure}

We start by investigating the throughput of the systems.
For this, data is fed into Kafka at the maximum sustainable rate for each configuration.
The throughput is the time difference between the first and the last processed input tuple divided by the number of input tuples.
We use the five queries shown in Figure~\ref{fig:queries_q5} where no additional filtering was imposed on the inputs and the full history of the input tuples is considered, and another test with ten queries, with additionally more partly overlapping joins.

In Figure~\ref{fig:throughput_q5_10} we see the throughput of these workloads, where Flink and Spark reach roughly the same performance as well as already a speed up of 1.4 with trivial sharing.
Flink's throughput is a smidge higher what can be explained with the overhead of our routing implementation.
Our approach of globally optimizing these queries brings us a speedup of 2.6 compared with the naive implementations.
The great potential of sharing state can be seen in Figure~\ref{fig:memory_q5_10} where we compare the Storm implementations of isolated and shared execution.
We see here, that with five queries running independently, 3.1 times the memory is required and with ten queries even 5.3 times.
For measuring the latency, we assign each tuple a timestamp when it arrives at the system and another timestamp when all join results with this tuple are computed, and record the differences between these timestamps.
Figure~\ref{fig:latency_q5_10} shows that the average latencies with shared multi query optimization are increased by 14 to 16\%, compared to the other modes.
This is due to the increased chance of selecting locally suboptimal probe orders which then leads to tuples taking longer in order to report a result.

\subsection{Impact of Adaptation to Individual Queries}

For this test we use a four-way linear join query of artificially generated relations $R(a), S(a,b), T(b,c), U(c)$ where the inputs arrive with a constant rate of 100k tuples per second.
The join attributes set such that each tuple will be part of one join result, i.e., half of the tuples find join partners during probing.
The window size is five seconds for each input and the epoch duration one second.
We initialize the optimizer with a little higher selectivity for $S(b), T(b)$ to make sure the probe orders $\langle S, R, T, U\rangle$ and $\langle T, U, R, S\rangle$ are selected.

\begin{figure}[tb]
  \centering
  \begin{subfigure}[b]{0.49\columnwidth}
    \resizebox{\columnwidth}{!}{
      \begin{tikzpicture}
        \begin{axis}[
            xlabel={time},
            ymin=0,
            ymax=100,
            ylabel={Latency [ms]},
            legend style={anchor=north east, at={(1,1)}, inner xsep=1pt, inner ysep=1pt},
            every axis plot/.append style={thick},
          ]
          \addplot[color=green, mark=square]
          coordinates {  (1, 57) (2, 56) (3, 57) (4, 57) (5, 56) (6, 56) (7, 57) (8, 55) (9, 56) (10, 53) (11, 55) (12, 56) (13, 56) (14, 53) (15, 57) (16, 70) (17, 72) (18, 72) (19, 65) (20, 62) (21, 57) (22, 55) (23, 54) (24, 54) (25, 57) (26, 58) (27, 56) (28, 57) (29, 53) (30, 53) };
          \addplot[color=purple, mark=*]
          coordinates { (1, 56) (2, 56) (3, 56) (4, 53) (5, 59) (6, 57) (7, 56) (8, 55) (9, 56) (10, 55) (11, 55) (12, 56) (13, 56) (14, 57) (15, 57) (16, 70) (17, 73) (18, 76) (19, 81) (20,95) };
          \legend{A, S}
        \end{axis}
      \end{tikzpicture}
    }
    \caption{A sudden increase in join selectivity renders a static join strategy unviable.}\label{fig:latency1}
  \end{subfigure}
  \begin{subfigure}[b]{0.49\columnwidth}
    \resizebox{\columnwidth}{!}{
      \begin{tikzpicture}
        \begin{axis}[
            xlabel={time},
            ymin=0,
            ymax=100,
            ylabel={Latency [ms]},
            legend style={anchor=north east, at={(1,1)}, inner xsep=1pt, inner ysep=1pt},
            every axis plot/.append style={thick},
          ]
          \addplot[color=green, mark=square]
          coordinates { (1, 57) (2, 56) (3, 54) (4, 58) (5, 56) (6, 56) (7, 56) (8, 56) (9, 57) (10, 57) (11, 55) (12, 53) (13, 56) (14, 57) (15, 56) (16, 57) (17, 53) (18, 50) (19, 48) (20, 43) (21, 39) (22, 34) (23, 35) (24, 36) (25, 36) (26, 35) (27, 35) (28, 35) };
          \addplot[color=purple, mark=*]
          coordinates { (1, 57) (2, 56) (3, 55) (4, 56) (5, 53) (6, 56) (7, 57) (8, 57) (9, 53) (10, 57) (11, 55) (12, 56) (13, 54) (14, 58) (15, 54) (16, 56) (17, 53) (18, 57) (19, 53) (20, 58) (21, 56) (22, 56) (23, 57) (24, 56) (25, 56) (26, 57) (27, 56) (28, 57) };
          \legend{A, S}
        \end{axis}
      \end{tikzpicture}
    }
    \caption{Adaptive join processing lowers the average end-to-end latency.}\label{fig:latency2}
  \end{subfigure}
  \caption{Adaptive execution.}\label{fig:adaptive_experiments}
\end{figure}
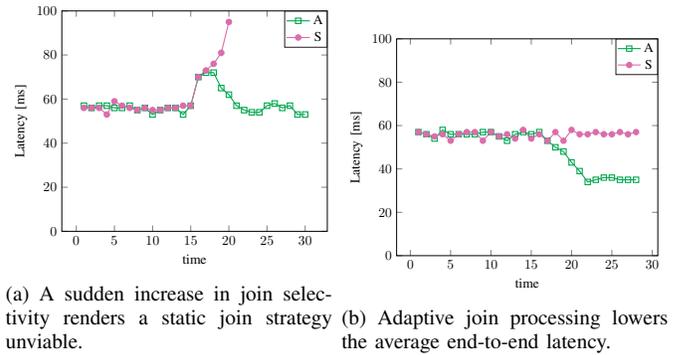

We compare the latencies of adaptive reoptimization {\bf (A)} and the initial static plans {\bf (S)} and initially they perform where similar with a little short of $56$ms latency, as depicted in Figure~\ref{fig:latency1}.
After 15 seconds the input changes drastically, now every tuple of $S$ finds 100 join partners in $R$, but none in $T$; vice versa for $T$-tuples.
Immediately after this the latency of both topologies increases slowly to about $72$ms, which is due to tuples being longer in buffers as the workers try to catch up.
In the adaptive strategy this works and after roughly a window a healthy latency is regained.
The static strategy cannot recover from this change and eventually the workers failed due to memory overflow.

We then use the same query, but with different input rates.
$R$ has 5M tuples per second, the other ones several orders of magnitude slower at 5k tuples per second.
In Figure~\ref{fig:latency2} we show the latency for the static topology which remains at the same constant level.
Again after 15 seconds we induce a change of the incoming data, now the size of the intermediate result of $S, T$, and $U$ gets very low.
This is recognized again after one epoch and a store for the result of the join of $S, T$, and $U$ is introduced.
We see a decline of the average latency that stabilizes from second 22 on a value of about 36ms.
During the decline phase, join partners are found already in the new store, but also older join partners need to be probed iteratively.

\subsection{ILP Optimization}

We simulate an environment consisting of multiple relations that can be joined together with given input rates and join selectivities.
In this environment we randomly generate queries and for each query we generate all probe orders and the corresponding ILP model as described in Section~\ref{sec:join_optimization}.
This model is solved using Gurobi.
We compare the cost of the joint query plan where query plans are shared and the cost for plainly applying only the best probe order for each query individually.
Tests were conducted on a system with 3.1\,GHz Intel Core i7 CPU and 16\,GB main memory.

The input relations have all the same arrival rate and a join between any two relations has a selectivity of arrival rate$^{-1}$.

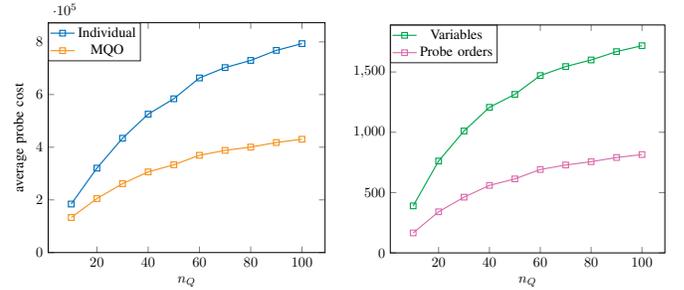
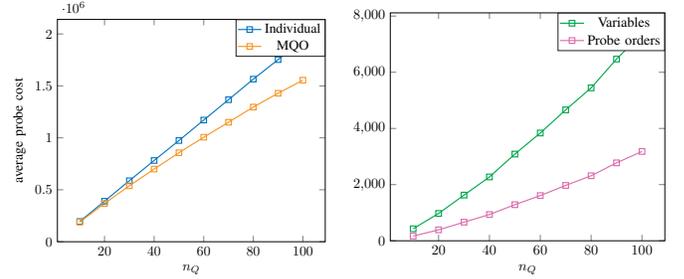
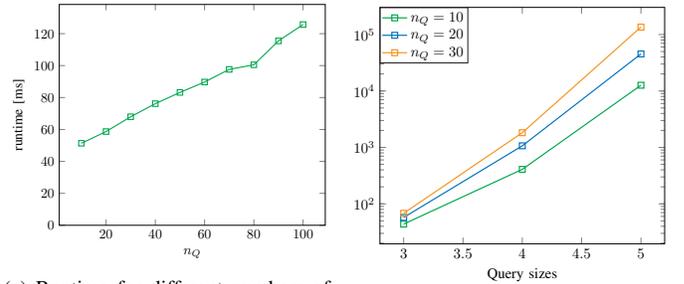
\begin{figure}[tb]
  \centering
  \begin{subfigure}[b]{0.49\columnwidth}
    \resizebox{\columnwidth}{!}{
      \begin{tikzpicture}
        \begin{axis}[
            xlabel={$n_Q$},
            ylabel={average probe cost},
            ymin=0,
            legend style={anchor=north west, at={(0,1)}, inner xsep=1pt, inner ysep=1pt},
            every axis plot/.append style={thick},
          ]
          \addplot[color=blue, mark=square]
          coordinates { (10,184260) (20,320600) (30,434120) (40,525580.0) (50,583920.0) (60,663100.0) (70,702340.0) (80,729640.0) (90,767740.0) (100,793340.0) };
          \addplot[color=orange, mark=square]
          coordinates { (10, 133110) (20, 204900) (30, 261520) (40, 306280.0) (50, 333370.0) (60, 369350.0) (70, 387990.0) (80, 400490.0) (90, 417890.0) (100, 430390.0) };
          \legend{Individual,MQO}
        \end{axis}
      \end{tikzpicture}
    }
    \caption{Probe cost of queries over three relations, drawing from 10 input relations.\\}\label{fig:subexp1}
  \end{subfigure}
  \hfill
  \begin{subfigure}[b]{0.49\columnwidth}
    \resizebox{\columnwidth}{!}{
      \begin{tikzpicture}
        \begin{axis}[
            xlabel={$n_Q$},
            ymin=0,
            legend style={anchor=north west, at={(0,1)}, inner xsep=1pt, inner ysep=1pt},
            every axis plot/.append style={thick},
          ]
          \addplot[color=green, mark=square]
          coordinates { (10, 389.76) (20, 761.56) (30, 1010.49) (40, 1206.73) (50, 1314.36) (60, 1470.24) (70, 1545.06) (80, 1599.52) (90, 1668.74) (100, 1717.8) };
          \addplot[color=purple, mark=square]
          coordinates { (10, 165.65) (20, 340.69) (30, 461.89) (40, 559.41) (50, 613.46) (60, 691.27) (70, 728.75) (80, 756.08) (90, 790.69) (100, 815.18) };
          \legend{Variables, Probe orders}
        \end{axis}
      \end{tikzpicture}
    }
    \caption{Problem sizes of queries over three relations, drawing from 10 input relations.\\}\label{fig:subexp2}
  \end{subfigure}

  \begin{subfigure}[b]{0.49\columnwidth}
    \resizebox{\columnwidth}{!}{
      \begin{tikzpicture}
        \begin{axis}[
            xlabel={$n_Q$},
            ylabel={average probe cost},
            ymin=0,
            legend style={anchor=north east, at={(1,1)}, inner xsep=1pt, inner ysep=1pt},
            every axis plot/.append style={thick},
          ]
          \addplot[color=blue, mark=square]
          coordinates { (10, 194740.00) (20, 390080.12) (30, 586920.18) (40, 782110.17) (50, 974180.06) (60, 1170980.36) (70, 1367020.18) (80, 1565020.52) (90, 1752870.71) (100, 1945310.47) };
          \addplot[color=orange, mark=square]
          coordinates { (10, 189340.00) (20, 368280.12) (30, 538170.18) (40, 699360.17) (50, 856580.06) (60, 1005730.36) (70, 1150070.18) (80, 1296070.52) (90, 1429070.71) (100, 1554760.47) };
          \legend{Individual,MQO}
        \end{axis}
      \end{tikzpicture}
    }
    \caption{Probe cost of queries over three relations, drawing from 100 input relations.\\}\label{fig:subexp3}
  \end{subfigure}
  \hfill
  \begin{subfigure}[b]{0.49\columnwidth}
    \resizebox{\columnwidth}{!}{
      \begin{tikzpicture}
        \begin{axis}[
            xlabel={$n_Q$},
            ymin=0,
            legend style={anchor=north east, at={(1,1)}, inner xsep=1pt, inner ysep=1pt},
            every axis plot/.append style={thick},
          ]
          \addplot[color=green, mark=square]
          coordinates { (10, 428.03) (20, 973.74) (30, 1626.08) (40, 2273.02) (50, 3086.7) (60, 3836.92) (70, 4661.48) (80, 5443.85) (90, 6469.75) (100, 7376.79) };
          \addplot[color=purple, mark=square]
          coordinates { (10, 167.36) (20, 389.46) (30, 662.65) (40, 934.8) (50, 1284.93) (60, 1610.21) (70, 1971.4) (80, 2314.98) (90, 2775.1) (100, 3180.1) };
          \legend{Variables, Probe orders}
        \end{axis}
      \end{tikzpicture}
    }
    \caption{Problem sizes of queries over three relations, drawing from 100 input relations.\\}\label{fig:subexp4}
  \end{subfigure}
  \begin{subfigure}[b]{0.49\columnwidth}
    \resizebox{\columnwidth}{!}{
      \begin{tikzpicture}
        \begin{axis}[
            xlabel={$n_Q$},
            ylabel={runtime [ms]},
            ymin=0,
            legend style={anchor=north east, at={(1,1)}, inner xsep=1pt, inner ysep=1pt},
            every axis plot/.append style={thick},
          ]
          \addplot[color=green, mark=square]
          coordinates { (10, 51.35) (20, 58.69) (30, 67.97) (40, 76.15) (50, 83.19) (60, 89.71) (70, 97.62) (80, 100.49) (90, 115.47) (100, 125.71) };
        \end{axis}
      \end{tikzpicture}
    }
    \caption{Runtime for different numbers of queries, drawing from 100 input relations.\\}\label{fig:subexp5}
  \end{subfigure}
  \hfill
  \begin{subfigure}[b]{0.49\columnwidth}
    \resizebox{\columnwidth}{!}{
      \begin{tikzpicture}
        \begin{axis}[
            xlabel={Query sizes},
            ymode=log,
            ymin=0,
            legend style={anchor=north west, at={(0,1)}, inner xsep=1pt, inner ysep=1pt},
            every axis plot/.append style={thick},
          ]
          \addplot[color=green, mark=square]
          coordinates { (3, 43.82) (4, 408.43) (5, 12639.16) };
          \addplot[color=blue, mark=square]
          coordinates { (3, 57.37) (4, 1067.08) (5, 45311.18)  };
          \addplot[color=orange, mark=square]
          coordinates { (3, 67.97) (4, 1822.86) (5, 135484.08) };
          \legend{$n_Q=10$, $n_Q=20$, $n_Q=30$}
        \end{axis}
      \end{tikzpicture}
    }
    \caption{Runtime for different query sizes, drawing from 100 input relations.\\}\label{fig:subexp6}
  \end{subfigure}
  \caption{ILP Experiments.}\label{fig:ilp_experiments}
\end{figure}

The first trial consists of ten input relations with three attributes each.
We generate $n_Q$ queries that each span three relations and eliminate exact duplicates (as these would be anyway answered together by a naive implementation).
Figure~\ref{fig:subexp1} shows the cost without sharing in the line for {\bf individual} optimization and with sharing for {\bf multi query optimization}.
The more queries we generate (over the same set of input relations), the higher the average probe cost gets for both.
But in case of multi query optimization, the probe cost of the MQO is significantly lower, around 50\%, than without sharing of probe order prefixes.

In Figure~\ref{fig:subexp2}, we show how the problem sizes grow: the number of variables fed into the ILP solver, indicated by the green line, grows more slowly the higher the number of queries; for 100 queries with each 3 relations, it is in average 1717.
This slow growth is because the more queries are optimized simultaneously, the more potential for sharing probe order prefixes there is, and each shared probe order prefix also shares a variable.
The purple line indicates the number of probe orders and it also grows slowly.
This is due to the fact that as we draw more queries over the small amount of input queries, the chances of producing the same query again increases.

We now examine the benefits for a higher number of input relations: the queries are now randomly drawn from 100 input relations, each with three attributes.
Figure~\ref{fig:subexp3} shows the probe cost savings, and here we see that for few queries nearly no savings are visible. For example, at 50 queries around 15\% of the cost can be saved.
In Figure~\ref{fig:subexp4} we see how also the problem size behaves more linearly.
If we look at the absolute numbers, we see, for example for $n_Q=50$ that 3000 variables are required compared to less than half of it in Figure~\ref{fig:subexp2}.
This is due to the fact the generated queries have very little overlap and thus only little possibility of sharing.
Both graphs are not linear but slightly convex.
This is because each new query also adds more possibilities for partitioning of a store, and each partitioning choice also increases the numbers of probe orders generated for a single query and consequently also the number of variables generated in the ILP.

In Figure~\ref{fig:subexp5}, we show the runtime for optimizing a different number of queries generated over 100 input relations, and see that it grows linearly, while even at 100 simultaneous queries, the optimization time is at 120 milliseconds.
In this experiment, all queries where over three relations.
We wanted to find out, how this approach scales to bigger queries, and thus altered the query size, i.e., the number of relations input into a query.
In Figure~\ref{fig:subexp6}, we see how the size of input relations effects optimization time.
Already ten queries of size four take 400ms---one order of magnitude more than ten queries of size three.
Optimizing ten queries of size five takes twelve seconds, and optimizing 30 queries of size five takes over two minutes.
While this adds to the delay of restructuring the topology to run in an optimized way, query answering can begin earlier with locally optimized probe orders defined on the input relations.

\subsection{Lessons Learned}

\begin{itemize}
  \item We have seen that there can be significant performance gains in combining multiple queries into one big topology. The increase in throughput means, that more tuples can be executed in the same time, thus the same cluster can process streams with a higher incoming rate.
  \item Sharing of state between operations increases the number of simultaneously answerable queries.
  \item Static joining ordering, like used in all currently available streaming systems, is prone to changes in the size of intermediate join results. A strategy for adopting to such changes avoids crashes, expensive recovery, or missing results.
  \item The optimization takes least time if the individual queries are smaller. Up to 30 queries of size five can be optimized within a second, which is still very usable for streaming scenarios.
\end{itemize}


\section{Conclusion}
\label{sec:conclusion}

We presented an approach for multi-query optimization of multi-way joins in scale-out streaming environments.
The speciality of the approach presented is a seamlessly integrated multi-stage optimization methodology,
starting from optimizing intra-operator multiway tuple routing, optimized join plans, and inter query optimization
by identifying common subplans. The multi query optimization is solved by generating and solving an integer linear program.
We detailed on the translation of the optimization result into a deployable operator topology and the requirements of the runtime such that the correct join result is produced. While the translation and implementation was tailored to Apache Storm, we are positive that the same strategy can be also applied to other stream processors, e.g., Flink using Stateful Functions.
Experiments showed that shared execution delivers a significant performance gain in terms of throughput and required memory.

\balance

\bibliographystyle{IEEEtran}
\bibliography{custom,main}

\begin{thebibliography}{10}
\providecommand{\url}[1]{#1}
\csname url@samestyle\endcsname
\providecommand{\newblock}{\relax}
\providecommand{\bibinfo}[2]{#2}
\providecommand{\BIBentrySTDinterwordspacing}{\spaceskip=0pt\relax}
\providecommand{\BIBentryALTinterwordstretchfactor}{4}
\providecommand{\BIBentryALTinterwordspacing}{\spaceskip=\fontdimen2\font plus
\BIBentryALTinterwordstretchfactor\fontdimen3\font minus
  \fontdimen4\font\relax}
\providecommand{\BIBforeignlanguage}[2]{{%
\expandafter\ifx\csname l@#1\endcsname\relax
\typeout{** WARNING: IEEEtran.bst: No hyphenation pattern has been}%
\typeout{** loaded for the language `#1'. Using the pattern for}%
\typeout{** the default language instead.}%
\else
\language=\csname l@#1\endcsname
\fi
#2}}
\providecommand{\BIBdecl}{\relax}
\BIBdecl

\bibitem{website:spark}
{Apache Software Foundation}, ``Apache {S}park,''
  \url{https://spark.apache.org/}, 2018.

\bibitem{DBLP:journals/debu/CarboneKEMHT15}
\BIBentryALTinterwordspacing
P.~Carbone, A.~Katsifodimos, S.~Ewen, V.~Markl, S.~Haridi, and K.~Tzoumas,
  ``Apache flink{\texttrademark}: Stream and batch processing in a single
  engine,'' \emph{{IEEE} Data Eng. Bull.}, vol.~38, no.~4, pp. 28--38, 2015.
  [Online]. Available: \url{http://sites.computer.org/debull/A15dec/p28.pdf}
\BIBentrySTDinterwordspacing

\bibitem{website:flink}
{Apache Software Foundation}, ``Apache {F}link,''
  \url{https://flink.apache.org/}, 2018.

\bibitem{website:storm}
``Apache {S}torm,'' \url{http://storm.apache.org/}, 2016.

\bibitem{DBLP:journals/pvldb/WangKSPZNRKS15}
\BIBentryALTinterwordspacing
G.~Wang, J.~Koshy, S.~Subramanian, K.~Paramasivam, M.~Zadeh, N.~Narkhede,
  J.~Rao, J.~Kreps, and J.~Stein, ``Building a replicated logging system with
  apache kafka,'' \emph{Proc. {VLDB} Endow.}, vol.~8, no.~12, pp. 1654--1655,
  2015. [Online]. Available:
  \url{http://www.vldb.org/pvldb/vol8/p1654-wang.pdf}
\BIBentrySTDinterwordspacing

\bibitem{website:kafka}
{Apache Software Foundation}, ``Apache {K}afka,''
  \url{https://kafka.apache.org/}, 2018.

\bibitem{DBLP:conf/sigmod/AnanthanarayananBDGJQRRSV13}
\BIBentryALTinterwordspacing
R.~Ananthanarayanan, V.~Basker, S.~Das, A.~Gupta, H.~Jiang, T.~Qiu,
  A.~Reznichenko, D.~Ryabkov, M.~Singh, and S.~Venkataraman, ``Photon:
  fault-tolerant and scalable joining of continuous data streams,'' in
  \emph{Proceedings of the {ACM} {SIGMOD} International Conference on
  Management of Data, {SIGMOD} 2013, New York, NY, USA, June 22-27, 2013},
  K.~A. Ross, D.~Srivastava, and D.~Papadias, Eds.\hskip 1em plus 0.5em minus
  0.4em\relax {ACM}, 2013, pp. 577--588. [Online]. Available:
  \url{http://doi.acm.org/10.1145/2463676.2465272}
\BIBentrySTDinterwordspacing

\bibitem{DBLP:journals/pvldb/KolchinskyS18}
\BIBentryALTinterwordspacing
I.~Kolchinsky and A.~Schuster, ``Join query optimization techniques for complex
  event processing applications,'' \emph{{PVLDB}}, vol.~11, no.~11, pp.
  1332--1345, 2018. [Online]. Available:
  \url{http://www.vldb.org/pvldb/vol11/p1332-kolchinsky.pdf}
\BIBentrySTDinterwordspacing

\bibitem{DBLP:conf/vldb/HammadFAE03}
\BIBentryALTinterwordspacing
M.~A. Hammad, M.~J. Franklin, W.~G. Aref, and A.~K. Elmagarmid, ``Scheduling
  for shared window joins over data streams,'' in \emph{Proceedings of 29th
  International Conference on Very Large Data Bases, {VLDB} 2003, Berlin,
  Germany, September 9-12, 2003}, J.~C. Freytag, P.~C. Lockemann, S.~Abiteboul,
  M.~J. Carey, P.~G. Selinger, and A.~Heuer, Eds.\hskip 1em plus 0.5em minus
  0.4em\relax Morgan Kaufmann, 2003, pp. 297--308. [Online]. Available:
  \url{http://www.vldb.org/conf/2003/papers/S10P02.pdf}
\BIBentrySTDinterwordspacing

\bibitem{DBLP:conf/dasfaa/YangZWX19}
\BIBentryALTinterwordspacing
J.~Yang, Y.~Zhang, J.~Wang, and C.~Xing, ``Distributed query engine for
  multiple-query optimization over data stream,'' in \emph{Database Systems for
  Advanced Applications - 24th International Conference, {DASFAA} 2019, Chiang
  Mai, Thailand, April 22-25, 2019, Proceedings, Part III, and {DASFAA} 2019
  International Workshops: BDMS, BDQM, and GDMA, Chiang Mai, Thailand, April
  22-25, 2019, Proceedings}, ser. Lecture Notes in Computer Science, G.~Li,
  J.~Yang, J.~Gama, J.~Natwichai, and Y.~Tong, Eds., vol. 11448.\hskip 1em plus
  0.5em minus 0.4em\relax Springer, 2019, pp. 523--527. [Online]. Available:
  \url{https://doi.org/10.1007/978-3-030-18590-9\_79}
\BIBentrySTDinterwordspacing

\bibitem{DBLP:conf/sigmod/KarimovRM19}
\BIBentryALTinterwordspacing
J.~Karimov, T.~Rabl, and V.~Markl, ``Astream: Ad-hoc shared stream
  processing,'' in \emph{Proceedings of the 2019 International Conference on
  Management of Data, {SIGMOD} Conference 2019, Amsterdam, The Netherlands,
  June 30 - July 5, 2019}, P.~A. Boncz, S.~Manegold, A.~Ailamaki, A.~Deshpande,
  and T.~Kraska, Eds.\hskip 1em plus 0.5em minus 0.4em\relax {ACM}, 2019, pp.
  607--622. [Online]. Available: \url{https://doi.org/10.1145/3299869.3319884}
\BIBentrySTDinterwordspacing

\bibitem{DBLP:conf/sigmod/DossingerMR19}
\BIBentryALTinterwordspacing
M.~Dossinger, S.~Michel, and C.~Roudsarabi, ``{CLASH:} {A} high-level
  abstraction for optimized, multi-way stream joins over apache storm,'' in
  \emph{Proceedings of the 2019 International Conference on Management of Data,
  {SIGMOD} Conference 2019, Amsterdam, The Netherlands, June 30 - July 5,
  2019}, P.~A. Boncz, S.~Manegold, A.~Ailamaki, A.~Deshpande, and T.~Kraska,
  Eds.\hskip 1em plus 0.5em minus 0.4em\relax {ACM}, 2019, pp. 1897--1900.
  [Online]. Available: \url{https://doi.org/10.1145/3299869.3320217}
\BIBentrySTDinterwordspacing

\bibitem{DBLP:conf/sigmod/ToshniwalTSRPKJGFDBMR14}
\BIBentryALTinterwordspacing
A.~Toshniwal, S.~Taneja, A.~Shukla, K.~Ramasamy, J.~M. Patel, S.~Kulkarni,
  J.~Jackson, K.~Gade, M.~Fu, J.~Donham, N.~Bhagat, S.~Mittal, and D.~V.
  Ryaboy, ``Storm@twitter,'' in \emph{International Conference on Management of
  Data, {SIGMOD} 2014, Snowbird, UT, USA, June 22-27, 2014}, C.~E. Dyreson,
  F.~Li, and M.~T. {\"{O}}zsu, Eds.\hskip 1em plus 0.5em minus 0.4em\relax
  {ACM}, 2014, pp. 147--156. [Online]. Available:
  \url{http://doi.acm.org/10.1145/2588555.2595641}
\BIBentrySTDinterwordspacing

\bibitem{DBLP:journals/ngc/KitsuregawaTM83}
\BIBentryALTinterwordspacing
M.~Kitsuregawa, H.~Tanaka, and T.~Moto{-}Oka, ``Application of hash to data
  base machine and its architecture,'' \emph{New Gener. Comput.}, vol.~1,
  no.~1, pp. 63--74, 1983. [Online]. Available:
  \url{https://doi.org/10.1007/BF03037022}
\BIBentrySTDinterwordspacing

\bibitem{DBLP:conf/sigmod/DeWittKOSSW84}
\BIBentryALTinterwordspacing
D.~J. DeWitt, R.~H. Katz, F.~Olken, L.~D. Shapiro, M.~Stonebraker, and D.~A.
  Wood, ``Implementation techniques for main memory database systems,'' in
  \emph{SIGMOD'84, Proceedings of Annual Meeting, Boston, Massachusetts, USA,
  June 18-21, 1984}, B.~Yormark, Ed.\hskip 1em plus 0.5em minus 0.4em\relax
  {ACM} Press, 1984, pp. 1--8. [Online]. Available:
  \url{https://doi.org/10.1145/602259.602261}
\BIBentrySTDinterwordspacing

\bibitem{DBLP:conf/sigmod/SelingerACLP79}
\BIBentryALTinterwordspacing
P.~G. Selinger, M.~M. Astrahan, D.~D. Chamberlin, R.~A. Lorie, and T.~G. Price,
  ``Access path selection in a relational database management system,'' in
  \emph{Proceedings of the 1979 {ACM} {SIGMOD} International Conference on
  Management of Data, Boston, Massachusetts, USA, May 30 - June 1}, P.~A.
  Bernstein, Ed.\hskip 1em plus 0.5em minus 0.4em\relax {ACM}, 1979, pp.
  23--34. [Online]. Available: \url{https://doi.org/10.1145/582095.582099}
\BIBentrySTDinterwordspacing

\bibitem{DBLP:journals/vldb/SteinbrunnMK97}
\BIBentryALTinterwordspacing
M.~Steinbrunn, G.~Moerkotte, and A.~Kemper, ``Heuristic and randomized
  optimization for the join ordering problem,'' \emph{{VLDB} J.}, vol.~6,
  no.~3, pp. 191--208, 1997. [Online]. Available:
  \url{https://doi.org/10.1007/s007780050040}
\BIBentrySTDinterwordspacing

\bibitem{DBLP:conf/sigmod/SchuhCD16}
\BIBentryALTinterwordspacing
S.~Schuh, X.~Chen, and J.~Dittrich, ``An experimental comparison of thirteen
  relational equi-joins in main memory,'' in \emph{Proceedings of the 2016
  International Conference on Management of Data, {SIGMOD} Conference 2016, San
  Francisco, CA, USA, June 26 - July 01, 2016}, F.~{\"{O}}zcan, G.~Koutrika,
  and S.~Madden, Eds.\hskip 1em plus 0.5em minus 0.4em\relax {ACM}, 2016, pp.
  1961--1976. [Online]. Available:
  \url{https://doi.org/10.1145/2882903.2882917}
\BIBentrySTDinterwordspacing

\bibitem{DBLP:journals/vldb/LeisRGMBKN18}
\BIBentryALTinterwordspacing
V.~Leis, B.~Radke, A.~Gubichev, A.~Mirchev, P.~A. Boncz, A.~Kemper, and
  T.~Neumann, ``Query optimization through the looking glass, and what we found
  running the join order benchmark,'' \emph{{VLDB} J.}, vol.~27, no.~5, pp.
  643--668, 2018. [Online]. Available:
  \url{https://doi.org/10.1007/s00778-017-0480-7}
\BIBentrySTDinterwordspacing

\bibitem{DBLP:conf/icdt/JoglekarR16}
\BIBentryALTinterwordspacing
M.~Joglekar and C.~R{\'{e}}, ``It's all a matter of degree: Using degree
  information to optimize multiway joins,'' in \emph{19th International
  Conference on Database Theory, {ICDT} 2016, Bordeaux, France, March 15-18,
  2016}, ser. LIPIcs, W.~Martens and T.~Zeume, Eds., vol.~48.\hskip 1em plus
  0.5em minus 0.4em\relax Schloss Dagstuhl - Leibniz-Zentrum fuer Informatik,
  2016, pp. 11:1--11:17. [Online]. Available:
  \url{https://doi.org/10.4230/LIPIcs.ICDT.2016.11}
\BIBentrySTDinterwordspacing

\bibitem{DBLP:journals/tods/KramerS09}
\BIBentryALTinterwordspacing
J.~Kr{\"{a}}mer and B.~Seeger, ``Semantics and implementation of continuous
  sliding window queries over data streams,'' \emph{{ACM} Trans. Database
  Syst.}, vol.~34, no.~1, pp. 4:1--4:49, 2009. [Online]. Available:
  \url{http://doi.acm.org/10.1145/1508857.1508861}
\BIBentrySTDinterwordspacing

\bibitem{DBLP:journals/vldb/HammadAE08}
\BIBentryALTinterwordspacing
M.~A. Hammad, W.~G. Aref, and A.~K. Elmagarmid, ``Query processing of multi-way
  stream window joins,'' \emph{{VLDB} J.}, vol.~17, no.~3, pp. 469--488, 2008.
  [Online]. Available: \url{https://doi.org/10.1007/s00778-006-0017-y}
\BIBentrySTDinterwordspacing

\bibitem{DBLP:conf/dasfaa/ZhouYYZ06}
\BIBentryALTinterwordspacing
Y.~Zhou, Y.~Yan, F.~Yu, and A.~Zhou, ``Pmjoin: Optimizing distributed multi-way
  stream joins by stream partitioning,'' in \emph{Database Systems for Advanced
  Applications, 11th International Conference, {DASFAA} 2006, Singapore, April
  12-15, 2006, Proceedings}, ser. Lecture Notes in Computer Science, M.~Lee,
  K.~Tan, and V.~Wuwongse, Eds., vol. 3882.\hskip 1em plus 0.5em minus
  0.4em\relax Springer, 2006, pp. 325--341. [Online]. Available:
  \url{https://doi.org/10.1007/11733836\_24}
\BIBentrySTDinterwordspacing

\bibitem{DBLP:conf/icdt/AfratiJRSU17}
\BIBentryALTinterwordspacing
F.~N. Afrati, M.~R. Joglekar, C.~R{\'{e}}, S.~Salihoglu, and J.~D. Ullman,
  ``{GYM:} {A} multiround distributed join algorithm,'' in \emph{20th
  International Conference on Database Theory, {ICDT} 2017, March 21-24, 2017,
  Venice, Italy}, ser. LIPIcs, M.~Benedikt and G.~Orsi, Eds., vol.~68.\hskip
  1em plus 0.5em minus 0.4em\relax Schloss Dagstuhl - Leibniz-Zentrum fuer
  Informatik, 2017, pp. 4:1--4:18. [Online]. Available:
  \url{https://doi.org/10.4230/LIPIcs.ICDT.2017.4}
\BIBentrySTDinterwordspacing

\bibitem{DBLP:conf/sigmod/ViglasN02}
\BIBentryALTinterwordspacing
S.~Viglas and J.~F. Naughton, ``Rate-based query optimization for streaming
  information sources,'' in \emph{Proceedings of the 2002 {ACM} {SIGMOD}
  International Conference on Management of Data, Madison, Wisconsin, USA, June
  3-6, 2002}, M.~J. Franklin, B.~Moon, and A.~Ailamaki, Eds.\hskip 1em plus
  0.5em minus 0.4em\relax {ACM}, 2002, pp. 37--48. [Online]. Available:
  \url{http://doi.acm.org/10.1145/564691.564697}
\BIBentrySTDinterwordspacing

\bibitem{DBLP:conf/vldb/ViglasNB03}
\BIBentryALTinterwordspacing
S.~Viglas, J.~F. Naughton, and J.~Burger, ``Maximizing the output rate of
  multi-way join queries over streaming information sources,'' in \emph{{VLDB}
  2003, Proceedings of 29th International Conference on Very Large Data Bases,
  September 9-12, 2003, Berlin, Germany}, J.~C. Freytag, P.~C. Lockemann,
  S.~Abiteboul, M.~J. Carey, P.~G. Selinger, and A.~Heuer, Eds.\hskip 1em plus
  0.5em minus 0.4em\relax Morgan Kaufmann, 2003, pp. 285--296. [Online].
  Available: \url{http://www.vldb.org/conf/2003/papers/S10P01.pdf}
\BIBentrySTDinterwordspacing

\bibitem{DBLP:conf/vldb/GolabO03}
\BIBentryALTinterwordspacing
L.~Golab and M.~T. {\"{O}}zsu, ``Processing sliding window multi-joins in
  continuous queries over data streams,'' in \emph{{VLDB} 2003, Proceedings of
  29th International Conference on Very Large Data Bases, September 9-12, 2003,
  Berlin, Germany}, J.~C. Freytag, P.~C. Lockemann, S.~Abiteboul, M.~J. Carey,
  P.~G. Selinger, and A.~Heuer, Eds.\hskip 1em plus 0.5em minus 0.4em\relax
  Morgan Kaufmann, 2003, pp. 500--511. [Online]. Available:
  \url{http://www.vldb.org/conf/2003/papers/S16P01.pdf}
\BIBentrySTDinterwordspacing

\bibitem{DBLP:conf/edbt/WangR09}
\BIBentryALTinterwordspacing
S.~Wang and E.~A. Rundensteiner, ``Scalable stream join processing with
  expensive predicates: workload distribution and adaptation by time-slicing,''
  in \emph{{EDBT} 2009, 12th International Conference on Extending Database
  Technology, Saint Petersburg, Russia, March 24-26, 2009, Proceedings}, ser.
  {ACM} International Conference Proceeding Series, M.~L. Kersten, B.~Novikov,
  J.~Teubner, V.~Polutin, and S.~Manegold, Eds., vol. 360.\hskip 1em plus 0.5em
  minus 0.4em\relax {ACM}, 2009, pp. 299--310. [Online]. Available:
  \url{http://doi.acm.org/10.1145/1516360.1516396}
\BIBentrySTDinterwordspacing

\bibitem{DBLP:conf/sigmod/LinOWY15}
\BIBentryALTinterwordspacing
Q.~Lin, B.~C. Ooi, Z.~Wang, and C.~Yu, ``Scalable distributed stream join
  processing,'' in \emph{Proceedings of the 2015 {ACM} {SIGMOD} International
  Conference on Management of Data, Melbourne, Victoria, Australia, May 31 -
  June 4, 2015}, T.~K. Sellis, S.~B. Davidson, and Z.~G. Ives, Eds.\hskip 1em
  plus 0.5em minus 0.4em\relax {ACM}, 2015, pp. 811--825. [Online]. Available:
  \url{http://doi.acm.org/10.1145/2723372.2746485}
\BIBentrySTDinterwordspacing

\bibitem{DBLP:journals/corr/NasirMGKS15a}
\BIBentryALTinterwordspacing
M.~A.~U. Nasir, G.~D.~F. Morales, D.~Garc{\'{\i}}a{-}Soriano, N.~Kourtellis,
  and M.~Serafini, ``Partial key grouping: Load-balanced partitioning of
  distributed streams,'' \emph{CoRR}, vol. abs/1510.07623, 2015. [Online].
  Available: \url{http://arxiv.org/abs/1510.07623}
\BIBentrySTDinterwordspacing

\bibitem{DBLP:conf/edbt/QiuPY19}
\BIBentryALTinterwordspacing
Y.~Qiu, S.~Papadias, and K.~Yi, ``Streaming hypercube: {A} massively parallel
  stream join algorithm,'' in \emph{Advances in Database Technology - 22nd
  International Conference on Extending Database Technology, {EDBT} 2019,
  Lisbon, Portugal, March 26-29, 2019}, M.~Herschel, H.~Galhardas, B.~Reinwald,
  I.~Fundulaki, C.~Binnig, and Z.~Kaoudi, Eds.\hskip 1em plus 0.5em minus
  0.4em\relax OpenProceedings.org, 2019, pp. 642--645. [Online]. Available:
  \url{https://doi.org/10.5441/002/edbt.2019.76}
\BIBentrySTDinterwordspacing

\bibitem{DBLP:conf/debs/MadsenZS16}
\BIBentryALTinterwordspacing
K.~G.~S. Madsen, Y.~Zhou, and L.~Su, ``Enorm: efficient window-based
  computation in large-scale distributed stream processing systems,'' in
  \emph{Proceedings of the 10th {ACM} International Conference on Distributed
  and Event-based Systems, {DEBS} '16, Irvine, CA, USA, June 20 - 24, 2016},
  A.~Gal, M.~Weidlich, V.~Kalogeraki, and N.~Venkasubramanian, Eds.\hskip 1em
  plus 0.5em minus 0.4em\relax {ACM}, 2016, pp. 37--48. [Online]. Available:
  \url{https://doi.org/10.1145/2933267.2933315}
\BIBentrySTDinterwordspacing

\bibitem{DBLP:conf/adbis/OguzYHED16}
\BIBentryALTinterwordspacing
D.~Oguz, S.~Yin, A.~Hameurlain, B.~Ergenc, and O.~Dikenelli, ``Adaptive join
  operator for federated queries over linked data endpoints,'' in
  \emph{Advances in Databases and Information Systems - 20th East European
  Conference, {ADBIS} 2016, Prague, Czech Republic, August 28-31, 2016,
  Proceedings}, ser. Lecture Notes in Computer Science, J.~Pokorn{\'{y}},
  M.~Ivanovic, B.~Thalheim, and P.~Saloun, Eds., vol. 9809.\hskip 1em plus
  0.5em minus 0.4em\relax Springer, 2016, pp. 275--290. [Online]. Available:
  \url{https://doi.org/10.1007/978-3-319-44039-2\_19}
\BIBentrySTDinterwordspacing

\bibitem{DBLP:conf/icde/RodigerIK016}
\BIBentryALTinterwordspacing
W.~R{\"{o}}diger, S.~Idicula, A.~Kemper, and T.~Neumann, ``Flow-join: Adaptive
  skew handling for distributed joins over high-speed networks,'' in \emph{32nd
  {IEEE} International Conference on Data Engineering, {ICDE} 2016, Helsinki,
  Finland, May 16-20, 2016}.\hskip 1em plus 0.5em minus 0.4em\relax {IEEE}
  Computer Society, 2016, pp. 1194--1205. [Online]. Available:
  \url{https://doi.org/10.1109/ICDE.2016.7498324}
\BIBentrySTDinterwordspacing

\bibitem{DBLP:conf/sigmod/LiRD18}
\BIBentryALTinterwordspacing
R.~Li, M.~Riedewald, and X.~Deng, ``Submodularity of distributed join
  computation,'' in \emph{Proceedings of the 2018 International Conference on
  Management of Data, {SIGMOD} Conference 2018, Houston, TX, USA, June 10-15,
  2018}, G.~Das, C.~M. Jermaine, and P.~A. Bernstein, Eds.\hskip 1em plus 0.5em
  minus 0.4em\relax {ACM}, 2018, pp. 1237--1252. [Online]. Available:
  \url{http://doi.acm.org/10.1145/3183713.3183728}
\BIBentrySTDinterwordspacing

\bibitem{DBLP:journals/inffus/GomesC08}
\BIBentryALTinterwordspacing
J.~S. Gomes and H.~Choi, ``Adaptive optimization of join trees for multi-join
  queries over sensor streams,'' \emph{Information Fusion}, vol.~9, no.~3, pp.
  412--424, 2008. [Online]. Available:
  \url{https://doi.org/10.1016/j.inffus.2007.06.001}
\BIBentrySTDinterwordspacing

\bibitem{DBLP:conf/icde/LiSMBCL07}
\BIBentryALTinterwordspacing
Q.~Li, M.~Shao, V.~Markl, K.~S. Beyer, L.~S. Colby, and G.~M. Lohman,
  ``Adaptively reordering joins during query execution,'' in \emph{Proceedings
  of the 23rd International Conference on Data Engineering, {ICDE} 2007, The
  Marmara Hotel, Istanbul, Turkey, April 15-20, 2007}, R.~Chirkova, A.~Dogac,
  M.~T. {\"{O}}zsu, and T.~K. Sellis, Eds.\hskip 1em plus 0.5em minus
  0.4em\relax {IEEE} Computer Society, 2007, pp. 26--35. [Online]. Available:
  \url{https://doi.org/10.1109/ICDE.2007.367848}
\BIBentrySTDinterwordspacing

\bibitem{DBLP:conf/sigmod/HellersteinA00}
\BIBentryALTinterwordspacing
R.~Avnur and J.~M. Hellerstein, ``Eddies: Continuously adaptive query
  processing,'' in \emph{Proceedings of the 2000 {ACM} {SIGMOD} International
  Conference on Management of Data, May 16-18, 2000, Dallas, Texas, {USA.}},
  W.~Chen, J.~F. Naughton, and P.~A. Bernstein, Eds.\hskip 1em plus 0.5em minus
  0.4em\relax {ACM}, 2000, pp. 261--272. [Online]. Available:
  \url{http://doi.acm.org/10.1145/342009.335420}
\BIBentrySTDinterwordspacing

\bibitem{DBLP:conf/vldb/TianD03}
\BIBentryALTinterwordspacing
F.~Tian and D.~J. DeWitt, ``Tuple routing strategies for distributed eddies,''
  in \emph{{VLDB} 2003, Proceedings of 29th International Conference on Very
  Large Data Bases, September 9-12, 2003, Berlin, Germany}, J.~C. Freytag,
  P.~C. Lockemann, S.~Abiteboul, M.~J. Carey, P.~G. Selinger, and A.~Heuer,
  Eds.\hskip 1em plus 0.5em minus 0.4em\relax Morgan Kaufmann, 2003, pp.
  333--344. [Online]. Available:
  \url{http://www.vldb.org/conf/2003/papers/S11P02.pdf}
\BIBentrySTDinterwordspacing

\bibitem{DBLP:conf/sigmod/Trummer017}
\BIBentryALTinterwordspacing
I.~Trummer and C.~Koch, ``Solving the join ordering problem via mixed integer
  linear programming,'' in \emph{Proceedings of the 2017 {ACM} International
  Conference on Management of Data, {SIGMOD} Conference 2017, Chicago, IL, USA,
  May 14-19, 2017}, S.~Salihoglu, W.~Zhou, R.~Chirkova, J.~Yang, and D.~Suciu,
  Eds.\hskip 1em plus 0.5em minus 0.4em\relax {ACM}, 2017, pp. 1025--1040.
  [Online]. Available: \url{https://doi.org/10.1145/3035918.3064039}
\BIBentrySTDinterwordspacing

\bibitem{DBLP:journals/pvldb/Trummer016}
\BIBentryALTinterwordspacing
------, ``Multiple query optimization on the d-wave 2x adiabatic quantum
  computer,'' \emph{{PVLDB}}, vol.~9, no.~9, pp. 648--659, 2016. [Online].
  Available: \url{http://www.vldb.org/pvldb/vol9/p648-trummer.pdf}
\BIBentrySTDinterwordspacing

\bibitem{DBLP:conf/iscis/DokerogluBC14}
\BIBentryALTinterwordspacing
T.~D{\"{o}}keroglu, M.~A. Bayir, and A.~Cosar, ``Integer linear programming
  solution for the multiple query optimization problem,'' in \emph{Information
  Sciences and Systems 2014 - Proceedings of the 29th International Symposium
  on Computer and Information Sciences, {ISCIS} 2014, Krakow, Poland, October
  27-28, 2014}, T.~Czach{\'{o}}rski, E.~Gelenbe, and R.~Lent, Eds.\hskip 1em
  plus 0.5em minus 0.4em\relax Springer, 2014, pp. 51--60. [Online]. Available:
  \url{https://doi.org/10.1007/978-3-319-09465-6\_6}
\BIBentrySTDinterwordspacing

\bibitem{DBLP:conf/cloud/JonathanCW18}
\BIBentryALTinterwordspacing
A.~Jonathan, A.~Chandra, and J.~B. Weissman, ``Multi-query optimization in
  wide-area streaming analytics,'' in \emph{Proceedings of the {ACM} Symposium
  on Cloud Computing, SoCC 2018, Carlsbad, CA, USA, October 11-13, 2018}.\hskip
  1em plus 0.5em minus 0.4em\relax {ACM}, 2018, pp. 412--425. [Online].
  Available: \url{https://doi.org/10.1145/3267809.3267842}
\BIBentrySTDinterwordspacing

\bibitem{DBLP:journals/dpd/WilschutA93}
\BIBentryALTinterwordspacing
A.~N. Wilschut and P.~M.~G. Apers, ``Dataflow query execution in a parallel
  main-memory environment,'' \emph{Distributed Parallel Databases}, vol.~1,
  no.~1, pp. 103--128, 1993. [Online]. Available:
  \url{https://doi.org/10.1007/BF01277522}
\BIBentrySTDinterwordspacing

\bibitem{DBLP:journals/pvldb/VitorovicEGMEDK16}
\BIBentryALTinterwordspacing
A.~Vitorovic, M.~Elseidy, K.~Guliyev, K.~V. Minh, D.~Espino, M.~Dashti,
  Y.~Klonatos, and C.~Koch, ``Squall: Scalable real-time analytics,''
  \emph{{PVLDB}}, vol.~9, no.~13, pp. 1553--1556, 2016. [Online]. Available:
  \url{http://www.vldb.org/pvldb/vol9/p1553-vitorovic.pdf}
\BIBentrySTDinterwordspacing

\bibitem{DBLP:conf/bigdataconf/Dossinger019}
\BIBentryALTinterwordspacing
M.~Dossinger and S.~Michel, ``Scaling out multi-way stream joins using
  optimized, iterative probing,'' in \emph{2019 {IEEE} International Conference
  on Big Data (Big Data), Los Angeles, CA, USA, December 9-12, 2019}.\hskip 1em
  plus 0.5em minus 0.4em\relax {IEEE}, 2019, pp. 449--456. [Online]. Available:
  \url{https://doi.org/10.1109/BigData47090.2019.9005973}
\BIBentrySTDinterwordspacing

\bibitem{DBLP:books/daglib/0090562}
A.~Schrijver, \emph{Theory of linear and integer programming}, ser.
  Wiley-Interscience series in discrete mathematics and optimization.\hskip 1em
  plus 0.5em minus 0.4em\relax Wiley, 1999.

\bibitem{website:tphc}
``The {TPC-H} benchmark,'' \url{http://www.tpc.org/tpch/}, 2016.

\end{thebibliography}

\end{document}